\documentclass{cernyrep}
\usepackage[T1]{fontenc}
\usepackage[bookmarks, colorlinks=true, linktoc=page, pdftex, linkcolor=black, citecolor=black, urlcolor=blue]{hyperref}
\usepackage{fancyhdr}
\fancyhfoffset{4 mm}
\fancypagestyle{ARTICLE}{%
\fancyhf{} 
\lhead{\hfill CERN Accelerator School Proceedings ---
{\it RF for Accelerators} ---  Berlin, Germany,  2023\hfill}
\lfoot{\hspace{3 mm} Available online at \url{https://cas.web.cern.ch/previous-schools}}
\rfoot{\thepage\hspace*{3mm}}

}
\fancypagestyle{ARTTITLE}{%
\fancyhf{}
\lhead{\hfill CERN Accelerator School Proceedings ---
{\it RF for Accelerators} ---  Berlin, Germany,  2023\hfill}
\lfoot{\hspace{3 mm} Available online at \url{https://cas.web.cern.ch/previous-schools}}
\rfoot{\thepage\hspace*{3mm}}

}

\usepackage{varwidth}
\usepackage{xcolor}
\usepackage{graphicx}
\usepackage{epsfig}
\usepackage{bm}
\usepackage{amsmath}
\usepackage{longtable}
\usepackage{mathtools}
\usepackage{siunitx}
\usepackage{appendix}
\usepackage{todonotes}
\graphicspath{{Figures/}}
\newcommand{\E}{\mathcal{E}}
\newcommand{\F}{\mathcal{F}}

\newcommand{\chk}[1]{#1}
\newcommand{\edt}[1]{#1}
\begin{document}

\title{Longitudinal beam dynamics II\edt{. Lecture 2}: intensity effects and beam instabilities}
\author{Elena Shaposhnikova\thanks{Elena.Chapochnikova@cern.ch}}

\date{\today}
\begin{abstract}
In this lecture, dealing with intensity effects, the equations of longitudinal motion with induced beam voltage are analyzed.
This allows \edt{us}, first, to describe \edt{the} single-bunch potential well distortion and \edt{the} loss of Landau damping. Then the linearized Vlasov equation is used to derive step-by-step the Lebedev equation.  This matrix equation is then applied to find the thresholds, growth rates and spectra of \edt{an} unstable multi-bunch beam. The lecture is concluded by \edt{a} discussion of possible beam-instability cures.
\end{abstract}

\maketitle
\thispagestyle{ARTTITLE}

\section{\label{sec:Introduction}Introduction}
\noindent
This lecture is a continuation of the previous one, where the equations of longitudinal motion in \edt{single and double rf systems} of a proton synchrotron were derived and analyzed without intensity effects included. Here we use the same notations and consider what happens when the voltage induced by the~beam in the accelerator environment is added to the rf voltage seen by the particles.
\edt{A general introduction to beam instabilities in accelerators can be found in the textbooks~\cite{Chao, NG}.} The material of this lecture is based on the papers~\cite{Ivan, Karpov2024}, where more details as well as \edt{a} more complete list of references can be found.

The lecture is organized in the following way. After this {\bf Introduction}, the main equations of longitudinal motion with intensity effects and solutions for a stationary particle distribution are presented in {\bf Section~\ref{sec:main_equations}}. The concepts of the beam-induced voltage and longitudinal impedance are introduced and then applied \edt{to the} description of the potential well distortion, namely, the synchronous phase shift and synchrotron frequency shift of a single bunch.
In {\bf Section~\ref{sec:perturbation}}, the Lebedev matrix equation is derived step-by-step from the linearized Vlasov equation using the perturbation formalism.
This general equation is then used  in {\bf Section~\ref{sec:Thesholds}} to obtain an analytic expression for the threshold of the loss of Landau damping (LLD) in the presence of \edt{a} constant inductive impedance.
In {\bf Section~\ref{sec:multibunch}}, the same equation is applied to a~multi-bunch beam in the presence of a narrow-band impedance. The thresholds, growth rates and spectra of the coupled-bunch instability, as well as the Robinson instability, are derived and it is shown how the threshold is modified by the broad-band impedance.
Possible cures of the longitudinal beam instabilities with some examples from the CERN accelerators are discussed in {\bf Section~\ref{sec:cures}}.
The main conclusions are presented in {\bf Section~\ref{sec:conclusions}}.

\section{\label{sec:main_equations} Longitudinal motion with intensity effects }
\subsection{Equations of motion}
We start with a stationary situation, while time-dependent perturbations are treated in the following section. We choose the coordinate system with respect to the synchronous particle with the energy $E_0$ and the phase $\phi_{s0}$ that corresponds to the case of a single rf system without taking into account intensity effects. Then, the main equations of longitudinal particle motion can be written using the conjugate variables \{$\Delta E/(h\omega_0)$, $\phi$\},
\begin{flalign}
    \frac{d\phi}{dt}= \dot{\phi} &= \frac{h^2\omega^2_0\eta}{\beta^2 E_0}\left(\frac{\Delta E}{h\omega_0} \right)  \label{eq:first_eq_motion},\\
    \frac{d}{dt}\left(\frac{\Delta E}{h\omega_0} \right) &= \frac{1}{2\pi h}\left[qV_t(\phi) - \delta E_0\right], \label{eq:second_eq_motion}
\end{flalign}
where $\Delta E$ and $\phi$ are respectively the energy and phase deviations of the particle, $\omega_0 = 2\pi f_0$, $f_0$ is the revolution frequency, $\beta$ is the particle velocity normalized by the speed of light, $h$ is the harmonic number, $q$ is the \edt{electric} charge of the particle, $\delta E_0$ is the energy gain per turn of the synchronous particle, $\eta = 1/\gamma^2_\mathrm{tr} - 1/\gamma^2$ is the slip factor, $\gamma$ is the relativistic Lorentz factor, and $\gamma_\mathrm{tr}$ is the Lorentz factor at transition energy. These equations can be found in standard textbooks (e.g.~\cite{Chao, NG}).
Note that the first equation of motion (\ref{eq:first_eq_motion}) is not modified by intensity effects.

Below we mainly consider the case of a single rf system, while the derivations can be easily adapted to any combination of rf waves.
In the presence of intensity effects, the total voltage $V_t(\phi) = V_\mathrm{rf}(\phi) + V_\mathrm{ind}(\phi)$, in addition to the rf voltage $V_\mathrm{rf}$, contains contributions from the beam-induced fields described as the beam-induced voltage $V_\mathrm{ind}$.
Then, $V_\mathrm{rf}(\phi) = V_0\sin(\phi_{s0} + \phi)$, with $V_0$ being the rf voltage amplitude, and the synchronous phase $\phi_{s0} = \arcsin{(\delta E_0/qV_0)}$ for $\eta<0$ or $\phi_{s0} =\pi-\arcsin{(\delta E_0/qV_0)}$ for $\eta>0$.
The acceleration rate of the synchronous particle is defined by the magnetic ramp and therefore its energy gain per turn $\delta E_0 = qV_0\sin\phi_{s0}$ does not change in the presence of intensity effects. This is why in Eq.~(\ref{eq:second_eq_motion}) we keep $\delta E_0$ defined by the phase $\phi_{s0}$ without intensity effects. The new synchronous phase $\phi_s = \phi_{s0} +\Delta\phi_s$, with intensity effects, is then found from the condition
\begin{equation}
    V_0 \sin{\phi_{s0}} = V_0\sin(\phi_{s0} + \Delta \phi_s) + V_\mathrm{ind}(\Delta \phi_s),
    \label{eq:phis_int}
\end{equation}
where $\Delta\phi_s$ is the synchronous phase shift due to intensity effects.

Combining the two equations of motion, the phase equation with intensity effects can be written as
\begin{equation}
    \frac{d^2\phi}{dt^2} + \frac{\omega^2_{s0}}{V_0 \cos\phi_{s0}}
    \left[V_t(\phi) - V_0\sin{\phi_{s0}} \right] = 0,
    \label{eq:phase_eq_int}
\end{equation}
where $\omega_{s0} = 2\pi f_{s0}$ is the angular frequency of small-amplitude synchrotron oscillations in a bare rf potential with
\begin{equation}
    \label{eq:synchrotron_tune_0}
    \omega_{s0}^2 = -\frac{h \, \omega_0^2 \eta q V_0\cos{\phi_{s0}}}{2\pi\beta^2 E_0}.
\end{equation}

\subsection{Beam-induced voltage and longitudinal impedance}
The voltage induced by the beam in the accelerator ring is a periodic function of the azimuth $\theta$, so that in the rf phase, $V_\mathrm{ind}(\phi) = V_\mathrm{ind}(\phi +2\pi h)$, and it can be presented as \edt{a} Fourier series
\begin{equation}
     V_\mathrm{ind}(\phi) =  \sum_{k=-\infty}^{\infty} V_k e^{i \frac{k}{h} \phi}.
\end{equation}
Each voltage harmonic $V_k$ is related to the corresponding harmonic $I_k$ of the beam current $I(\phi)$ by the longitudinal impedance $Z_k = Z(k\omega_0)$ at frequency $k\omega_0$, $V_k = - Z_k I_k$ (see \edt{e.g. the} lecture on \edt{i}mpedances and wakefields by A.~Mostacci at this School). For a single bunch with $N_p$ particles the~beam current is $I(\phi) = q h\omega_0 N_p \lambda(\phi) = 2\pi h I_b \lambda(\phi)$, where $I_b = qN_p f_0$ is the average bunch current and
$\lambda(\phi) = (d N/d\phi)/N_p$ is the normalized line density. Then the stationary beam-induced voltage can be written in the form
\begin{equation}
    \label{eq:induced_voltage}
    V_\mathrm{ind}(\phi) =  \sum_{k=-\infty}^{\infty} V_k e^{i \frac{k}{h} \phi} = - q N_p \, h \, \omega_0 \sum_{k=-\infty}^{\infty} Z_k \lambda_k  e^{i \frac{k}{h} \phi} = -2\pi h I_b \sum_{k=-\infty}^{\infty} Z_k \lambda_k  e^{i \frac{k}{h} \phi},
\end{equation}
where $\lambda_k$ is the Fourier harmonic of the normalized line density
\begin{equation}
    \label{eq:line_density_harm}
    \lambda_k = \frac{1}{2\pi h} \int_{-\pi h}^{\pi h}d\phi \; \lambda(\phi) e^{-i \frac{k}{h} \phi}.
\end{equation}
The line density is defined by the distribution function $\F(\phi,\dot{\phi})$, i.e. $\lambda(\phi) = \int_{-\infty}^{\infty}d\dot{\phi}\, \F(\phi,\dot{\phi})$, and the~following normalization is imposed
\begin{equation}
    \label{eq:normalization}
    \int_{-\pi h}^{\pi h}d\phi\, \lambda(\phi) = \int_{-\pi h}^{\pi h}d\phi \int_{-\infty}^{\infty}d\dot{\phi}\, \F(\phi,\dot{\phi}) = 1.
\end{equation}

The longitudinal impedance $Z(\omega) = \mathrm{Re}Z(\omega) + i\,\mathrm{Im}Z(\omega)$ is a complex function of frequency. Its real (resistive) part is responsible for the beam energy loss, beam loading, beam-induced heating and beam instabilities, while the imaginary (reactive) part leads to the potential well distortion and loss of Landau damping. Many impedance sources in the accelerator ring can be modeled by \edt{a} resonator
\begin{equation}
    \label{eq:resonator}
    Z(\omega) = \frac{R_{sh}}{1 + iQ\left(\omega/\omega_r - \omega_r/\omega\right)},
\end{equation}
where $R_{sh}$ is the shunt impedance, $\omega_r$ is the resonant frequency and $Q$ is the quality factor, which defines the resonator bandwidth $\Delta\omega_r = \omega_r/(2Q)$. \edt{Note that this form of the resonator impedance corresponds to the time convention $e^{i\Omega t}$ used throughout this lecture; in the opposite convention $e^{-i\Omega t}$, adopted in some textbooks (e.g.~\cite{Chao}), the sign of the $iQ$ term is reversed.} Depending on the bandwidth, one distinguishes
\begin{itemize}
    \item narrow-band impedances (rf cavities and other cavity-like objects), for which the wake field induced by one bunch is still seen by the following bunches. These impedances are responsible for multi-bunch effects (beam loading, coupled-bunch instabilities);
    \item broad-band impedances (space charge, changes of the vacuum chamber cross-section, ...), with a~bandwidth larger than the inverse bunch length $\tau$, $\tau \Delta\omega_r > 1$, which are responsible for single-bunch effects (potential well distortion, loss of Landau damping, single-bunch instabilities).
\end{itemize}
For a multi-bunch beam, the effect of the voltage induced in the rf cavities is called beam loading (see \edt{also the} lecture by H.~Damerau at this School).

\begin{figure}[htb]
	\centering
	\includegraphics[width = 0.6\textwidth]{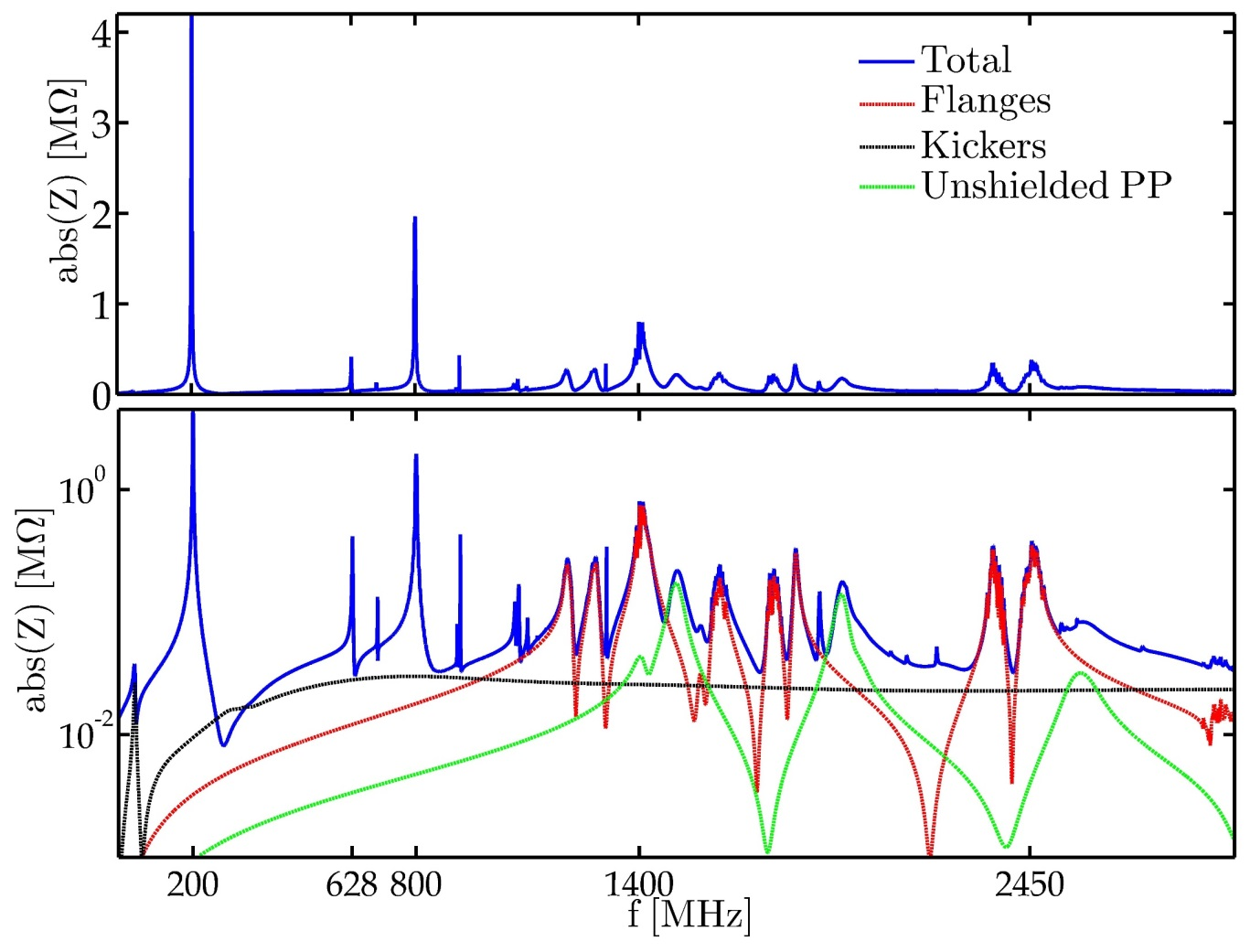}
	\caption{Longitudinal impedance model of the CERN SPS (before the impedance reduction during the long shutdown 2 in 2019--2020) in linear (top) and logarithmic (bottom) scales. The contributions of the vacuum flanges, kickers and unshielded pumping ports (PP) are shown separately. From Ref.~\cite{Lasheen2018}, see also~\cite{SPSimp}.}
	\label{fig:SPS_impedance}
\end{figure}

An example of a realistic longitudinal impedance model, developed for the CERN SPS, is shown in Fig.~\ref{fig:SPS_impedance}.

\begin{figure}[htb]
	\centering
	\begin{tabular}{cc}
	\includegraphics[width = 0.55\textwidth]{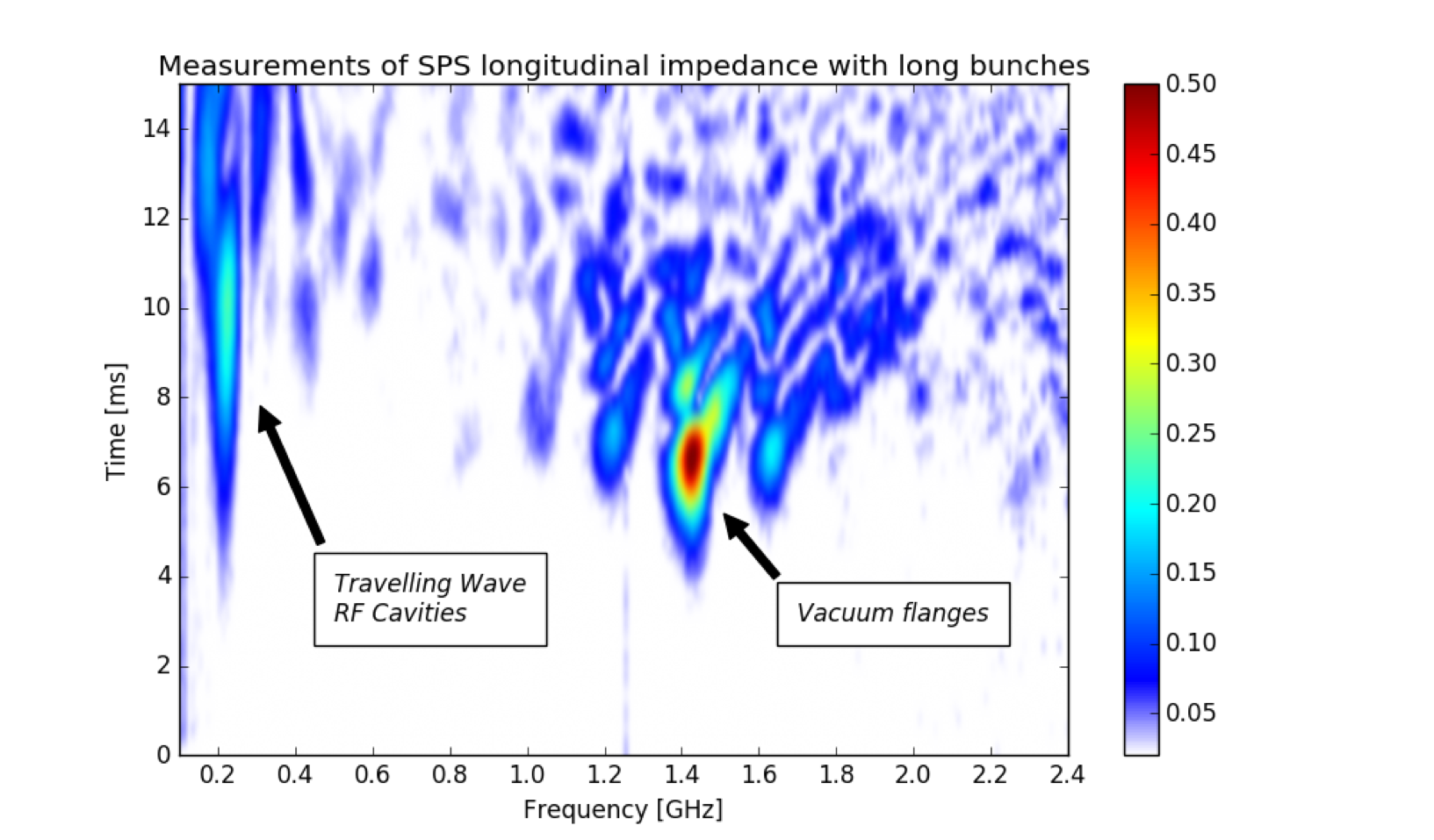} &
	\includegraphics[width = 0.40\textwidth]{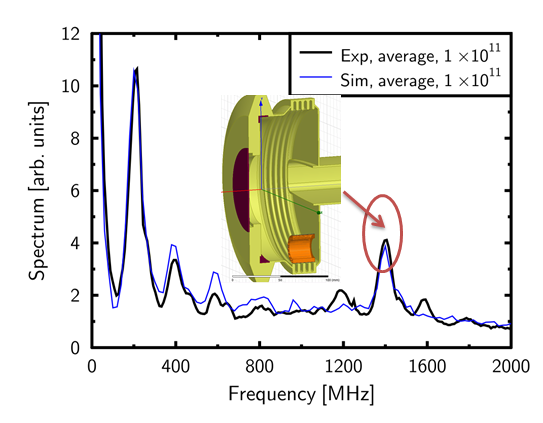}
	\end{tabular}
	\caption{Left: measured spectrum of unstable single bunches injected \edt{into} the CERN SPS at 26~GeV/c with the rf system switched off, as a function of time after injection. The line density modulation grows at the resonant frequencies of the travelling wave rf cavities and of the vacuum flanges. Right: averaged spectra of the unstable bunches from measurements and from macro-particle simulations with the SPS impedance model. The peak around 1.4~GHz corresponds to the impedance of the vacuum flanges (an example is shown in the \edt{inset}). From Ref.~\cite{Lasheen2018}.}
	\label{fig:unstable_spectrum}
\end{figure}

It contains contributions from the 200~MHz and 800~MHz travelling wave rf cavities, including their higher order modes (HOMs), from the kicker magnets, about 500 vacuum flanges, about 200 beam position monitors, unshielded pumping ports, beam scrapers, electrostatic septa, resistive wall and other elements.
Many impedance sources were identified and quantified in this ring using \edt{the} measurements of the spectrum of \edt{the} unstable single bunches injected \edt{into} the ring with the rf system switched off~\cite{Bohl1997}. In this case a slow debunching is accompanied by a fast instability with a line density modulation growing at the resonant frequencies of the impedances with high $R_{sh}/Q$, see Fig.~\ref{fig:unstable_spectrum}.

This method was used in the SPS as a tool for the two impedance reduction campaigns (see Section~\ref{sec:cures}).

\subsection{Potential well distortion}
The total voltage $V_t$ includes the voltage induced by the stationary line density $\lambda(\phi)$ and by its perturbation $\Tilde{\lambda}(\phi,t)$ (see the next Section). In the first case, the initial particle distribution is modified and a new equilibrium (stationary solution) is established. This effect is called the potential well distortion and it is usually considered as a single-bunch effect.

For small deviations $\Delta\phi = \phi - \phi_{s0}$ from the synchronous phase, $|\Delta\phi| \ll 1$, the phase equation (\ref{eq:phase_eq_int}) takes the form
\begin{equation}
    \frac{d^2\Delta\phi}{dt^2} + \omega^2_{s0}\left[\Delta\phi + \frac{V_\mathrm{ind}(\phi)}{V_0 \cos\phi_{s0}} \right] = 0.
    \label{eq:phase_eq_small}
\end{equation}
Using expression (\ref{eq:induced_voltage}) for the induced voltage in the case $k\Delta\phi/h \ll 1$ for all harmonics $k$ of the bunch spectrum, we have
\begin{equation}
    V_\mathrm{ind}(\phi) = -2\pi h I_b \sum_{k=-\infty}^{\infty} Z_k \lambda_k  e^{i \frac{k}{h} \Delta\phi} \simeq -2\pi h I_b \sum_{k=-\infty}^{\infty} Z_k \lambda_k  \left[1 + \frac{i k\Delta\phi}{h} + ...\right].
\end{equation}
For a symmetric bunch profile, the line density harmonics are real and $\lambda_k = \lambda_{-k}$. Taking into account that $\mathrm{Re}Z_k = \mathrm{Re}Z_{-k}$ and $\mathrm{Im}Z_k = -\mathrm{Im}Z_{-k}$, one obtains
\begin{equation}
    V_\mathrm{ind}(\phi) \simeq -2\pi h I_b \sum_{k=-\infty}^{\infty} \lambda_k  \left[\mathrm{Re}Z_k - \frac{ k\Delta\phi}{h}\mathrm{Im}Z_k + ...\right].
\end{equation}
After substitution of this expression into Eq.~(\ref{eq:phase_eq_small}), we can see that the first term defines the shift of the~synchronous phase, or the shift of the potential well,
\begin{equation}
    \Delta\phi_s \simeq \frac{2\pi h I_b}{V_0 \cos\phi_{s0}} \sum_{k=-\infty}^{\infty} \lambda_k  \mathrm{Re}Z_k,
    \label{eq:phase_shift}
\end{equation}
while the second term modifies the frequency of small-amplitude synchrotron oscillations,
\begin{equation}
    \omega^2_s \simeq \omega^2_{s0}\left[1 + \frac{2\pi I_b}{V_0 \cos\phi_{s0}} \sum_{k=-\infty}^{\infty} k \lambda_k  \mathrm{Im}Z_k \right].
    \label{eq:freq_shift}
\end{equation}

\subsubsection{Synchronous phase shift}
The synchronous phase shift (\ref{eq:phase_shift}) is defined by the resistive part of the impedance. It is \edt{only} valid for particles with small oscillation amplitude\edt{s}, but \edt{this shift} is difficult to measure. In practice, one measures the~shift of the bunch centre $\Delta\phi_b$ with respect to the rf voltage. It can be found from the~energy change $U$ per turn and per particle due to the induced voltage (negative for the energy loss), $U = q\int d\phi\, V_\mathrm{ind}(\phi)\lambda(\phi)$, which is compensated by the rf system, so that
\[
    V_0\sin\phi_{s0} \simeq V_0\sin\phi_{s0} + \Delta\phi_b \cos\phi_{s0}\, V_0 + U/q,
\]
and therefore
\begin{equation}
    \Delta\phi_b \simeq - \frac{U}{qV_0\cos\phi_{s0}}, \hspace{1 cm} U = - \chk{4\pi^2}\, q h^2 I_b \sum_{k=-\infty}^{\infty} |\lambda_k|^2  \mathrm{Re}Z_k.
    \label{eq:bunch_shift}
\end{equation}
The energy loss $|U|$ is proportional to the loss factor of the ring.
Since the bunch spectrum $\lambda_k$ depends on the bunch length, measurements of $\Delta\phi_b$ for various bunch lengths give an \edt{estimate} of the resistive impedance $\mathrm{Re}Z(\omega)$ of the ring.

\begin{figure}[htb]
	\centering
	\includegraphics[width = 0.42\textwidth]{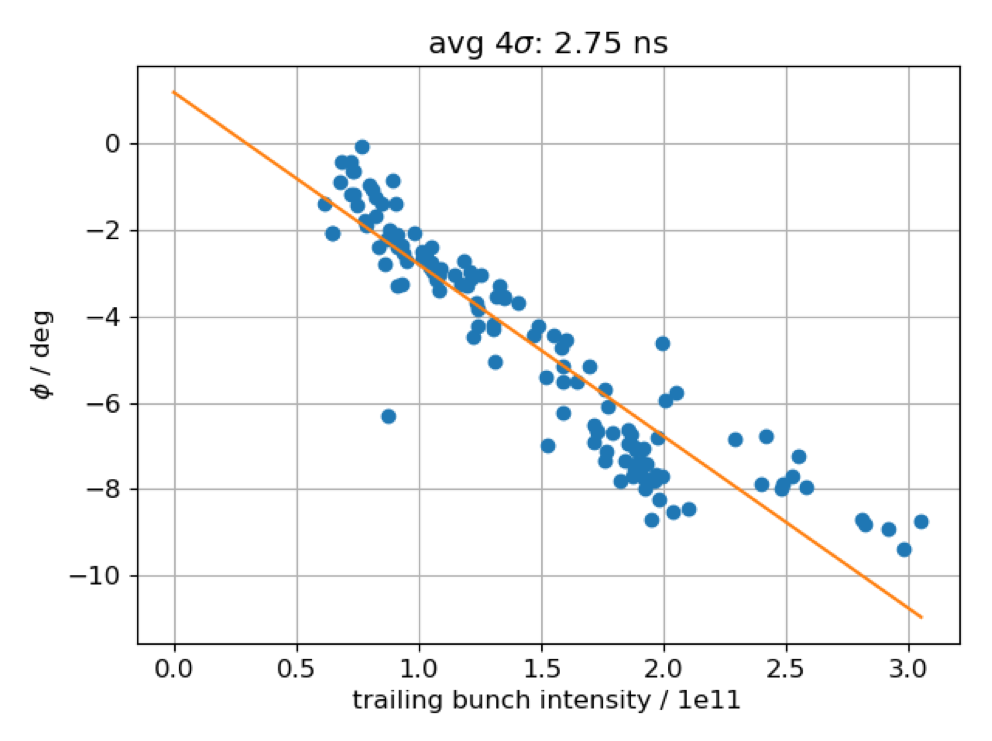}\\
	\begin{tabular}{cc}
	\includegraphics[width = 0.42\textwidth]{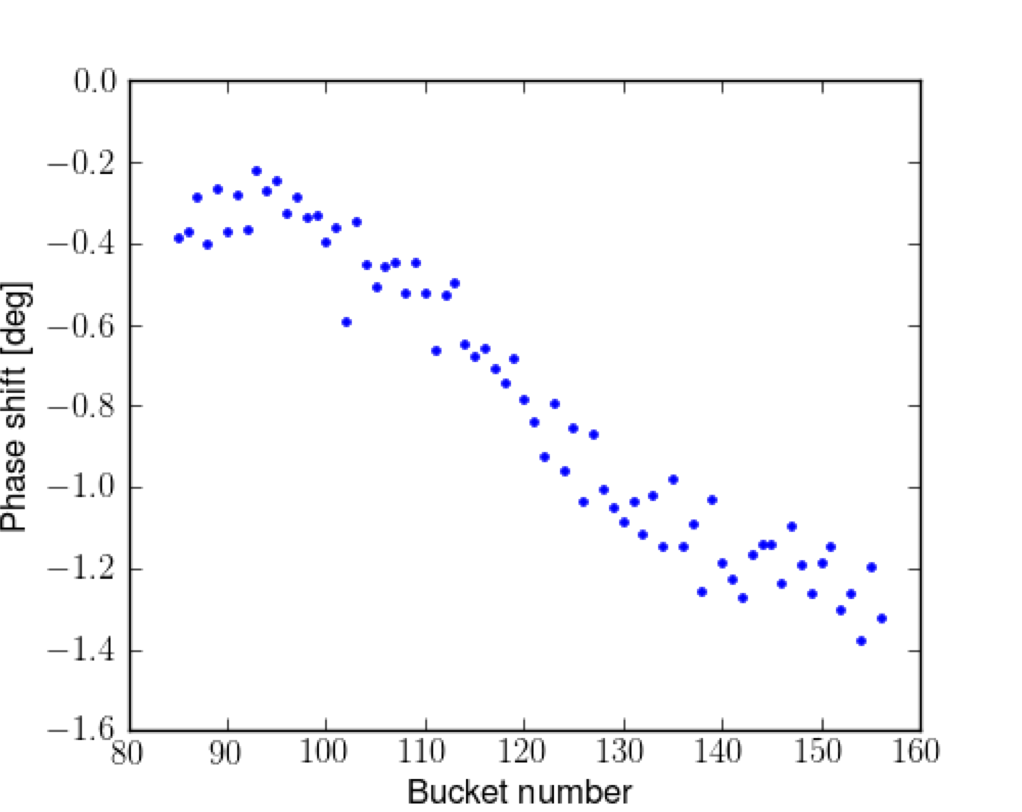} &
	\includegraphics[width = 0.45\textwidth]{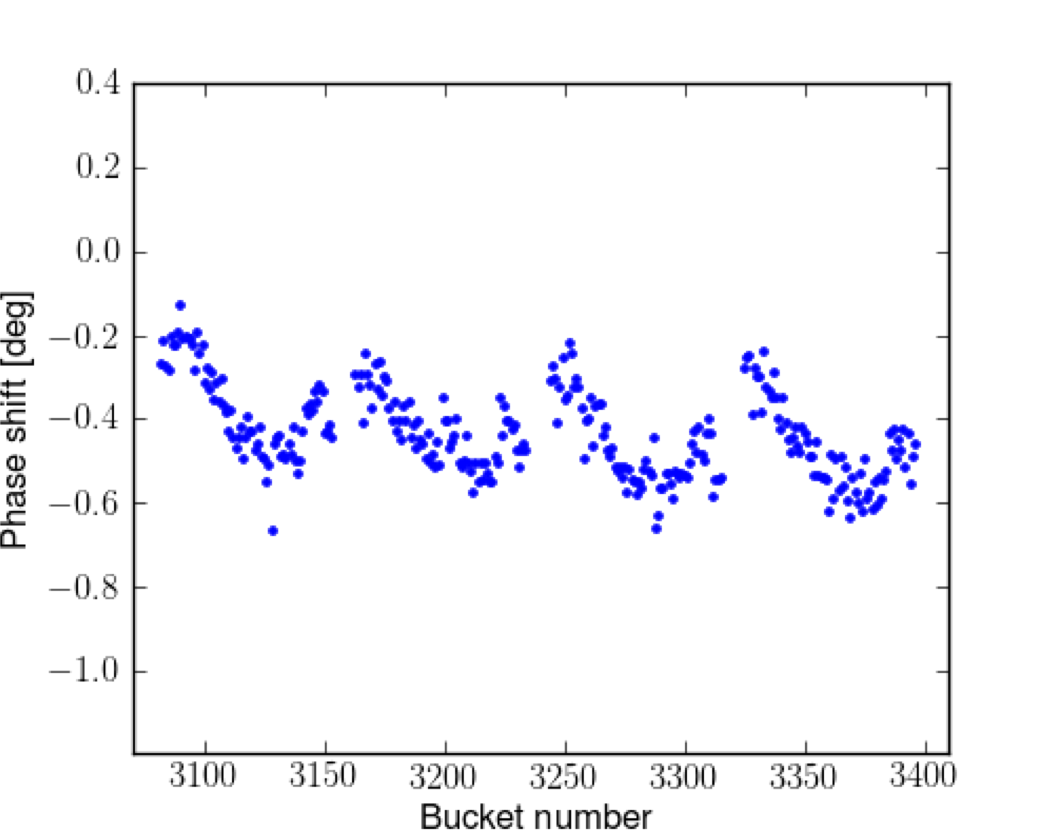}
	\end{tabular}
	\caption{Measurements of the synchronous phase shift. Top: \edt{the two-bunch method in the CERN SPS:} the phase of \edt{a} bunch of variable intensity measured with respect to \edt{a leading} bunch \edt{of} low intensity\edt{, for an average $4\sigma$ bunch length of 2.75~ns; from Ref.~\cite{Schwarz2020}}. Bottom: bunch-by-bunch phase shift \edt{measured in the LHC} along the bunch trains before (left) and after (right) the LHC surface scrubbing with beam, which reduced the electron cloud. From Refs.~\cite{EstebanMuller2015, JMPhD}.}
	\label{fig:phase_shift}
\end{figure}

\edt{The} synchronous phase shift \edt{can be} measured in two different ways: (1) as the phase variation between the beam signal and the rf signal\edt{, as it was done in the SPS~\cite{BohlEPAC04}}, and (2) as \edt{the} distance between two bunches\edt{, one} of low and \edt{one of} variable intensity \edt{(see Fig.~\ref{fig:phase_shift} (top) for an example of such a measurement in the CERN SPS~\cite{Schwarz2020})}.

In the \edt{first} case, using as a reference the rf signal from the power amplifier to the cavity, the energy loss due to beam loading is included in the measurements, while using the probe in the cavity, the beam loading is excluded. The measured bunch-by-bunch phase shift is proportional to the energy loss of each bunch and this is an operational tool in the LHC for the observation of the electron cloud, which builds up along the bunch train\edt{~\cite{EstebanMuller2015, JMPhD}}. An example in Fig.~\ref{fig:phase_shift} (bottom) shows how the \edt{phase shift due to the} electron cloud was reduced by the LHC surface scrubbing with beam.

\subsubsection{Synchrotron frequency shift}
\begin{figure}[htb]
	\centering
	\includegraphics[width = 0.6\textwidth]{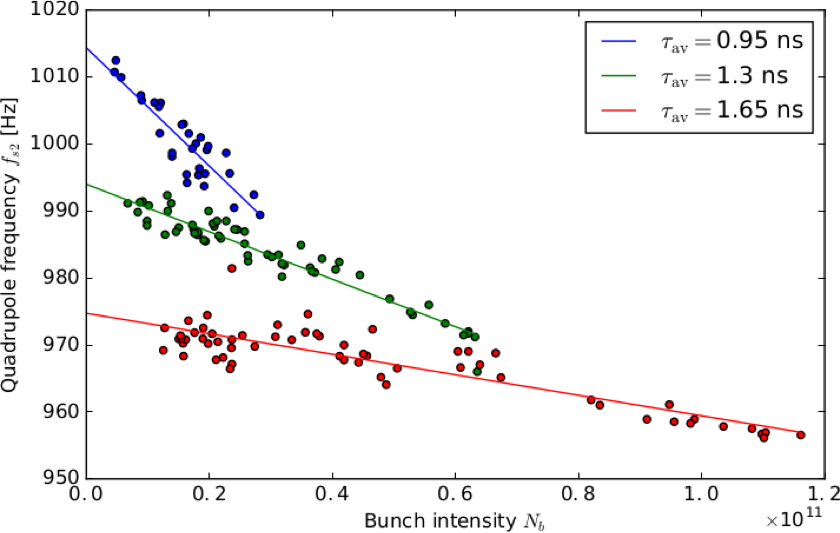}
	\caption{Quadrupole frequency $f_{s2}$ of the bunch-length oscillations measured in the CERN SPS at 26~GeV/c as a~function of bunch intensity for three different average bunch lengths $\tau_\mathrm{av}$, with the linear fits. The slope of the~intensity dependence gives an estimate of the reactive impedance $\mathrm{Im}Z/k$ of the ring. From Ref.~\cite{Lasheen2017}.}
	\label{fig:quad_freq}
\end{figure}

The synchrotron frequency shift (\ref{eq:freq_shift}) is defined by the reactive part of the impedance. For a small shift $\Delta\omega_s = \omega_s - \omega_{s0} \ll \omega_{s0}$ and \edt{an} impedance $\mathrm{Im}Z_k/k = \mathrm{Im}Z/k = $~const over the bunch spectrum, we have
\begin{equation}
    \Delta\omega_s \simeq \omega_{s0} \frac{\pi I_b}{V_0 \cos\phi_{s0}} \frac{\mathrm{Im}Z}{k} \sum_{k=-\infty}^{\infty} k^2 \lambda_k.
\end{equation}
Since $\lambda(\phi) = \sum_k \lambda_k e^{ik\phi/h}$, the last sum is defined by the curvature of the line density in the bunch centre
\[
    \sum_{k=-\infty}^{\infty} k^2 \lambda_k = - h^2 \left. \frac{d^2\lambda}{d\phi^2}\right|_{\phi = 0}.
\]
For example, for a parabolic bunch with the line density $\lambda(\phi) = \lambda_0 (1 - \phi^2/\phi_b^2)$ and the half bunch length $\phi_b = h\omega_0\tau/2$, where $\tau$ is the full bunch length in seconds, one obtains
\begin{equation}
    \omega^2_s = \omega^2_{s0}\left[1 + \frac{3 I_b}{\pi^2 V_0 h \cos\phi_{s0} (f_0\tau)^3} \frac{\mathrm{Im}Z}{k}  \right].
    \label{eq:freq_shift_parabolic}
\end{equation}
Note the strong dependence of the synchrotron frequency shift on the bunch length. Above transition energy ($\cos\phi_{s0}<0$) the inductive impedance ($\mathrm{Im}Z/k >0$) has a defocusing effect, i.e. the synchrotron frequency decreases with intensity and more rf voltage is needed to keep the same bunch length. The~intensity dependence of the synchrotron frequency (or of the quadrupole frequency $f_{s2} \simeq 2f_s$ of the~bunch-length oscillations after an rf mismatch at injection) measured for different bunch lengths can \edt{also be} used for \edt{the} evaluation of the reactive impedance $\mathrm{Im}Z/k$ of the ring~\cite{Lasheen2017}, see an example of such measurements in the CERN SPS in Fig.~\ref{fig:quad_freq}.

Note that for the same intensity the frequency shift is larger for shorter bunches, as expected from Eq.~(\ref{eq:freq_shift_parabolic}), while the frequency at zero intensity is lower for longer bunches due to the nonlinearity of the~rf voltage (see the previous lecture).

As follows from the analysis of the previous lecture, for short bunches the longitudinal emittance $\varepsilon_b \propto \omega_s \tau^2$. The reduction of the small-amplitude synchrotron frequency due to the potential well distortion leads to the bunch lengthening for a constant longitudinal emittance. In the short-bunch approximation, using Eq.~(\ref{eq:freq_shift_parabolic}), the bunch length $\tau$ with intensity effects is related to the zero-intensity bunch length $\tau_0$ as \edt{(see e.g.~\cite{Hofmann77, Hofmann1979})}
\begin{equation}
    1 = \left(\frac{\tau}{\tau_0}\right)^4 + \frac{\tau}{\tau_0} \, \frac{\omega_s^2(\tau_0) - \omega_{s0}^2}{\omega_{s0}^2}.
\end{equation}

\subsubsection{Stationary solution}
The first step in determining the instability threshold is to find a stationary particle distribution with the~voltage induced by this distribution. For lepton bunches, the energy spread is determined by the~synchrotron radiation and the line density is the solution of the Haissinski equation~\cite{Haissinski}. For proton (ion) bunches, the solution should be found for the distribution function.

It is convenient for further analysis to use another set of variables ($\mathcal{E}$, $\psi$),  which correspond respectively to the energy and phase of the synchrotron oscillations,
\begin{align}
    \mathcal{E} &= \frac{{\dot{\phi}}^2}{2\omega^2_{s0}} + U_t(\phi), \label{eq:energy} \\
    \psi &= \text{sgn}(\eta\Delta E)\frac{\omega_s(\mathcal{E})}{\sqrt{2}\omega_{s0}} \int_{\phi_\mathrm{max}}^\phi \frac{d\phi^\prime}{\sqrt{\mathcal{E} -U_t\left(\phi^\prime\right)}}.
\label{eq:phase}
\end{align}
The total potential can be obtained from the total voltage $V_t$
\begin{equation}
    \label{eq:potential}
    U_t(\phi) = \frac{1}{V_0\cos{\phi_{s0}}}\int^\phi_{\Delta \phi_{s}} d\phi^\prime \, \left[V_t(\phi^\prime) - V_0\sin{\phi_{s0}} \right]
\end{equation}
with the synchronous phase shift due to intensity effects $\Delta \phi_s$ that satisfies the relation (\ref{eq:phis_int}).
The dependence of the synchrotron frequency on the energy of synchrotron oscillations $\omega_s(\mathcal{E}) =2\pi/T_s(\mathcal{E})$ in Eq.~(\ref{eq:phase}) can be found from the period of oscillations
\begin{equation}
    \label{eq:synchrotron_freq}
    T_s(\mathcal{E})= \frac{\sqrt{2}}{\omega_{s0}} \int_{\chk{\phi_\mathrm{min}(\E)}}^{\phi_\mathrm{max}(\E)}\frac{d\phi^\prime}{ \sqrt{\mathcal{E} -U_t\left(\phi^\prime\right)} },
\end{equation}
where $\phi_\mathrm{min}(\mathcal{E})$ and $\phi_\mathrm{max}(\mathcal{E})$ are the minimum and maximum phases of the particle with the energy of synchrotron oscillation $\mathcal{E}$ which satisfy the relation $\mathcal{E}  = U_t(\phi)$. In these variables, the equations of unperturbed particle motion are simply
\begin{equation}
    \dot{\psi} = \omega_s(\E), \hspace{1 cm} \dot{\E} = 0.
    \label{eq:motion_E_psi}
\end{equation}

The stationary particle distribution function $\F$ is
a function of only the energy of synchrotron oscillations $\E$
with
the line density
\begin{equation}
    \label{eq:norm_line_density}
    \lambda(\phi) = 2\omega_{s0} \int_{U_t(\phi)}^{\E_{\max}} d\E \frac{\F(\E)}{\sqrt{2\left[\E - U_t(\phi) \right]}}.
\end{equation}
It depends on the total potential\edt{, which} in calculations can be found using an iterative procedure, similar to the one used in Ref.~\cite{burov2012van} (see Appendix~B in Ref.~\cite{Ivan}), assuming some form of the initial particle distribution. Below we will mostly consider distributions belonging to \edt{the} binomial family, which covers a wide range of realistic bunch shapes, from flat ($\mu = -1/2$) to Gaussian ($\mu \to \infty $),
\begin{equation}
    \label{eq:binomial_distribution}
    \F(\E) = \frac{1}{2\pi \omega_{s0} A_N}\left(1 - \frac{\E}{\E_\mathrm{max}}\right)^\mu = \frac{g\left(\E\right)}{2\pi \omega_{s0} A_N},
\end{equation}
where the normalization factor
\begin{equation}
    \label{eq:normalization_factor}
    A_N = \omega_{s0}\int_0^{\E_{\max}}d\E \,\frac{  g\left(\E\right)}{\omega_s(\E)}.
\end{equation}
In this case, the integration in Eq.~(\ref{eq:norm_line_density}) can be performed analytically yielding
\begin{equation}
    \label{eq:line_density_binom}
    \lambda(\phi) = \frac{\sqrt{ \E_{\max}} \Gamma(\mu+1)}{\sqrt{2\pi}A_N\Gamma(\mu+3/2)} \left[1-\frac{U_t(\phi)}{\E_{\max}}\right]^{\mu+1/2},
\end{equation}
where $\Gamma$ is the gamma function.

For $\mu \to \infty$ a bunch has a Gaussian line density and the corresponding bunch length $\tau_{4\sigma}$ is typically defined as four times the \edt{root-mean-square (RMS)} bunch length $\sigma$, i.e. $\tau_{4\sigma} = 4 \sigma$. The bunch length $\tau_{4\sigma}$ can be easily calculated from the \edt{full width at half maximum} (FWHM) bunch length $\tau_\mathrm{FWHM}$:
\begin{equation}
   \label{eq:bunch_length}
   \tau_{4\sigma} = \tau_\mathrm{FWHM} \sqrt{2/\ln 2}.
\end{equation}
In practice, the proton bunches are usually not Gaussian and have a finite full length $\tau_\mathrm{full}$ defined as
\begin{equation}
    \label{eq:full_bunch_length}
    \tau_\mathrm{full} = \left[\phi_{\max}(\E_{\max}) - \phi_{\min}(\E_{\max}) \right]/\omega_\mathrm{rf}.
\end{equation}
However, we will also use definition~(\ref{eq:bunch_length}) below as it has important features related to the LLD threshold (see \edt{Section}~\ref{sec:Thesholds}).

\section{\label{sec:perturbation} Perturbation formalism}
\subsection{Linearized Vlasov equation}
For \edt{the} analysis of beam stability,
we need to consider the perturbations $\Tilde{\F}$, $\Tilde{\lambda}$ and $\Tilde{V}_\mathrm{ind}$, respectively, to the~equilibrium distribution function $\F$, line density $\lambda$, and induced voltage $V_\mathrm{ind}$, introduced in the~previous section. If these deviations grow with time, the beam is unstable.

The Vlasov equation, which follows from the Liouville theorem $d\F/dt=0$, in variables ($\E,\psi$) is
\begin{equation}
    \label{eq:vlasov_equation_full}
    \frac{\partial \F}{\partial t} + \frac{d\E}{dt} \frac{\partial \F}{\partial\E} + \frac{d\psi}{dt} \frac{\partial \F}{\partial \psi} = 0,
\end{equation}
so that for a stationary case, according to Eqs.~(\ref{eq:motion_E_psi}), $\F = \F(\E)$.
Assuming the time dependence of the~perturbation in the form $\Tilde{\F}(\E,\psi,t) = \Tilde{\F}(\E,\psi,\Omega) e^{\chk{i\Omega t}}$, the beam is unstable if $\chk{\mathrm{Im}\,\Omega <0}$. Below we follow the notations and sign conventions of Ref.~\cite{Karpov2024}.
The linearized Vlasov equation in ($\E,\psi$) variables is
\begin{equation}
    \label{eq:vlasov_equation}
    \frac{\partial \Tilde{\F}}{\partial t} + \frac{d\E}{dt} \frac{d \F}{d\E} + \frac{d\psi}{dt} \frac{\partial \Tilde{\F}}{\partial \psi} = 0,
\end{equation}
where by definition $d\psi/dt =\omega_s(\E)$, while $d\E/dt$ can be found from a combination of the first and second equations of particle motion in the presence of perturbation
\begin{equation}
    \label{eq:second_eq_motion_perturbed}
    \frac{d\dot{\phi}}{dt} + \frac{\omega^2_{s0}}{V_0 \cos\phi_{s0}}
    \left[V_t(\phi) - V_0\sin{\phi_{s0}} \right] = - \frac{\omega^2_{s0}}{V_0 \cos\phi_{s0}} \Tilde{V}_\mathrm{ind}(\phi,t) .
\end{equation}
Multiplying both sides of Eq.~(\ref{eq:second_eq_motion_perturbed}) by $\dot{\phi}$
and taking into account Eq.~(\ref{eq:energy}) one obtains
\begin{equation}
    \label{eq:energy_derivative}
    \frac{d\E}{dt} = -\frac{d\phi}{dt}\frac{\Tilde{V}_\mathrm{ind}(\phi,t)}{V_0 \cos\phi_{s0}}  = -\omega_s(\E)\frac{\partial \Tilde{U}_\mathrm{ind}(\phi,t)}{\partial \psi},
\end{equation}
where the perturbed induced potential $\Tilde{U}_\mathrm{ind}(\phi,t)$ is defined similarly to Eq.~(\ref{eq:potential})
\begin{equation}
    \label{eq:induced_potential}
    \Tilde{U}_\mathrm{ind} (\phi) = \frac{1}{V_0\cos{\phi_{s0}}}\int^\phi_{\Delta \phi_{s}}d\phi^\prime \Tilde{V}_\mathrm{ind}(\phi^\prime).
\end{equation}
Finally, the linearized Vlasov equation can be written in the form
\begin{equation}
    \label{eq:vlasov_equation_new}
    \left[\frac{\partial }{\partial t} + \omega_s \frac{\partial }{\partial \psi}\right] \Tilde{\F} = \omega_s \frac{\partial \Tilde{U}_\mathrm{ind}}{\partial \psi} \frac{d \F}{d\E}.
\end{equation}

\subsection{Lebedev equation}
\edt{Here we derive a general solution of the linearized Vlasov equation~(\ref{eq:vlasov_equation_new}), which does not neglect the~synchrotron frequency spread: the matrix equation proposed by A.~N.~Lebedev in 1968~\cite{Lebedev1968}.} The detailed derivation with slightly different variables can \edt{also be} found in~\cite{Shaposhnikova1994}.
The solution of Eq.~(\ref{eq:vlasov_equation_new}) must be periodic in $\psi$, and can be represented as the sum of the harmonics $e^{-i m \psi}$ ($m\neq 0$).
Presenting the perturbations
for a given $\Omega$ as $\Tilde{\F} (\E,\psi,t) = \Tilde{\F} (\E,\psi,\Omega) e^{i\Omega t}$ and $\Tilde{U}_\mathrm{ind} (\E,\psi,t) = \Tilde{U}_\mathrm{ind} (\E,\psi,\Omega) e^{i\Omega t}$,
we have
\begin{equation}
    \Tilde{\F}(\E,\psi,\Omega) = \sum_{m = -\infty}^{\infty} \Tilde{\F}_m (\E,\Omega) e^{-i m \psi}, \;\;
    \Tilde{U}_\mathrm{ind} (\E,\psi,\Omega) = \sum_{m = -\infty}^{\infty} \Tilde{U}_{\mathrm{ind},m} (\E,\Omega) e^{-i m \psi},
\end{equation}
with
\begin{equation}
    \label{eq:harmonics_m}
    X_m (\E,\Omega) =\frac{1}{2\pi} \int_{-\pi}^{\pi}d\psi \, X(\E,\psi,\Omega) e^{i m\psi}.
\end{equation}
The azimuthal mode number $m = 1, 2, 3, ...$ defines the type of the bunch oscillations (dipole, quadrupole, sextupole, ...).
Then the solution of Eq.~(\ref{eq:vlasov_equation_new}) is
\begin{equation}
    \label{eq:solution}
    \Tilde{\F} (\E,\psi,\Omega) = - \omega_s (\E) \frac{d \F}{d\E} \sum_{m = -\infty}^{\infty}  \frac{m \Tilde{U}_{\mathrm{ind},m}(\E,\Omega) }{\Omega - m \omega_s(\E)}e^{-i m\psi}.
\end{equation}
The perturbed induced voltage is related to the perturbed line density as (see Appendix A in Ref.~\cite{Ivan})
\begin{equation}
    \label{eq:ind_vs_line}
    \Tilde{V}_k(\Omega) = - q N_p \, h\, \omega_0 Z_k(\Omega) \Tilde{\lambda}_k (\Omega),
\end{equation}
where $Z_k(\Omega) = Z (k\omega_0 + \Omega)$. Then, using Eq.~(\ref{eq:induced_potential}), the perturbed induced potential can be presented in the form
\begin{equation}
    \label{eq:ind_potential_perturb}
    \Tilde{U}_\mathrm{ind}(\E,\psi,\Omega) = \frac{i\zeta}{\cos\phi_{s0}}\sum_{k=-\infty}^{\infty}\frac{Z_k(\Omega)}{k} \Tilde{\lambda}_k(\Omega) \left( e^{i \frac{k}{h} \phi} - e^{i \frac{k}{h}\Delta \phi_{s}} \right),
\end{equation}
and its Fourier harmonics $\Tilde{U}_{\mathrm{ind},m}$, defined by Eq.~(\ref{eq:harmonics_m}), can be obtained as
\begin{equation}
    \label{eq:potential_hamonics_m}
    \Tilde{U}_{\mathrm{ind},m}(\E,\Omega) = \frac{i\zeta}{\cos\phi_{s0}} \sum_{k=-\infty}^{\infty}\frac{Z_k(\Omega)}{k} \Tilde{\lambda}_k(\Omega) I_{mk}(\E).
\end{equation}
Here we introduce the ``intensity'' parameter  $\zeta$ (in units of $\Omega^{-1}$),
\begin{equation}
    \label{eq:xi}
     \zeta = \frac{ q N_p \, h^2 \, \omega_0 }{V_0} = \frac{ 2\pi h^2 I_b }{V_0},
\end{equation}
so that the product $\zeta Z_k$ is dimensionless.

\begin{figure}[htb]
	\centering
	\begin{tabular}{cc}
	\includegraphics[width = 0.47\textwidth]{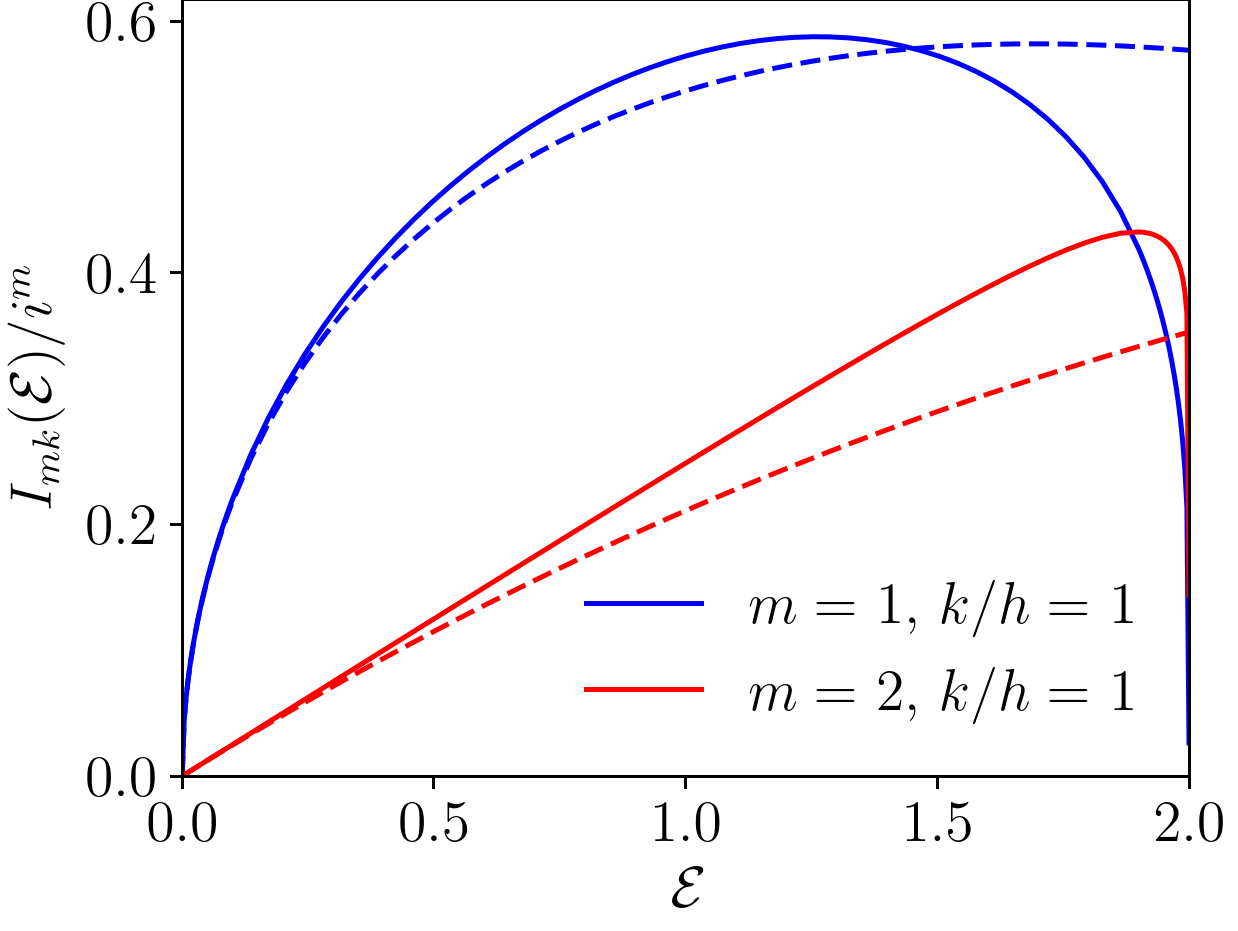} &
	\includegraphics[width = 0.47\textwidth]{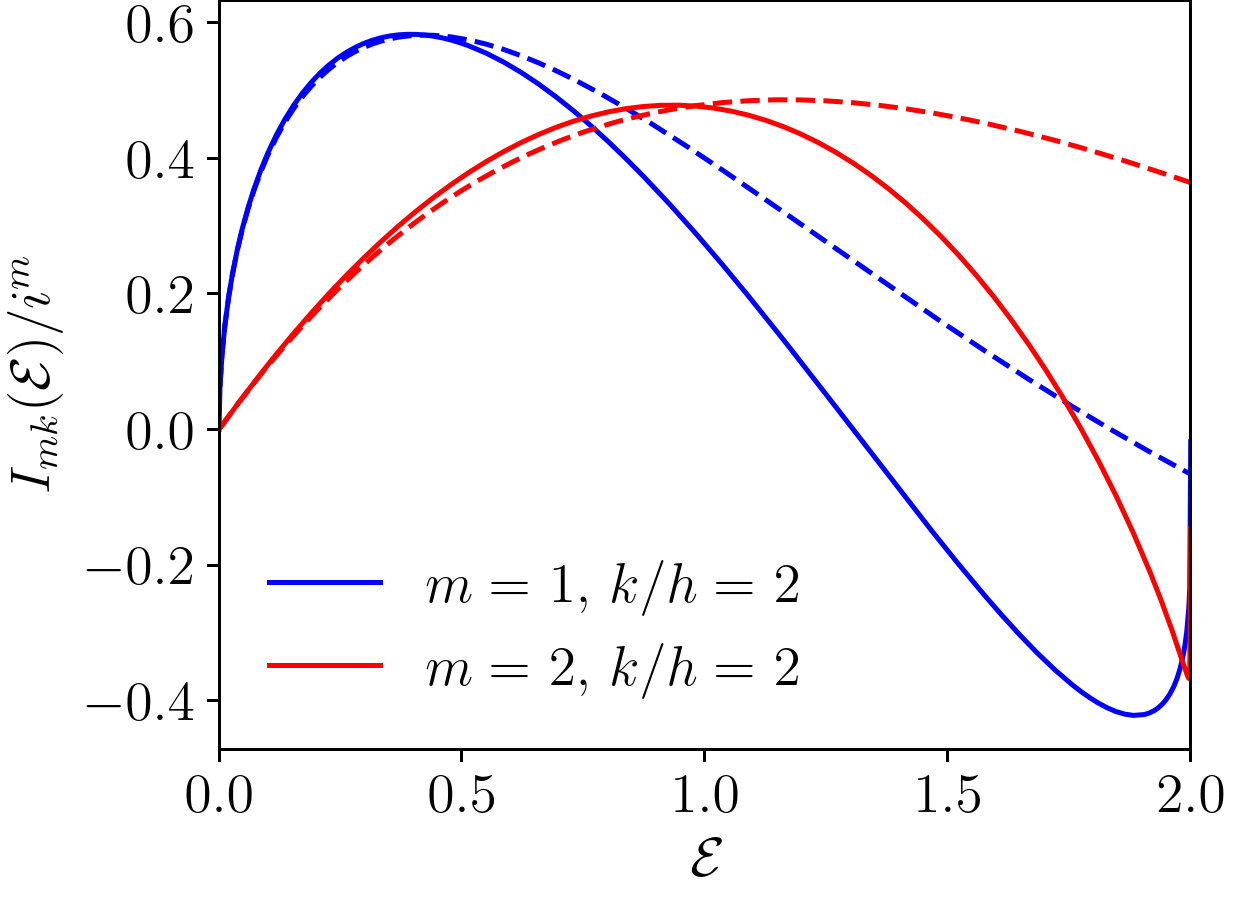}
	\end{tabular}
	\caption{Functions $I_{mk}(\E)/i^m$ in a single rf system without intensity effects for $k/h=1$ (left) and $k/h=2$ (right) calculated using the exact expression (\ref{eq:Imk}), solid lines, and its short-bunch approximation (\ref{eq:Imk_approx}), dashed lines, for the dipole ($m=1$) and quadrupole ($m=2$) modes.}
	\label{fig:Imk}
\end{figure}

The function $I_{mk}(\E)$, introduced for the first time in~\cite{Lebedev1968}, plays an important role\edt{:}
\begin{equation}
    \label{eq:Imk}
    I_{m k} (\E) = \frac{1}{2\pi} \int_{-\pi}^{\pi} d\psi \; e^{i\frac{k}{h}\phi(\E,\psi) + i m \psi} = \frac{1}{\pi} \int_{0}^{\pi} d\psi \; e^{i\frac{k}{h}\phi(\E,\psi)} \cos m \psi.
\end{equation}
These functions are the Fourier harmonics of the expansion of the azimuthal plane wave $e^{ik\phi/h}$ over the~harmonics $m\psi$ of the synchrotron motion.
The last expression
comes from the fact that $\phi(\E,-\psi) = \phi(\E,\psi)$. Note that $I_{mk}$ depends on $\zeta$ because $\phi(\E,\psi)$ is modified via potential well distortion. It is \edt{also} easy to see the following properties
\[
I_{-mk} = I_{mk},\; \text{and} \; I_{m-k} = I^*_{mk}.
\]
In the case of \edt{a} symmetric potential well, we also have $I^*_{mk} = (-1)^m I_{mk}$.
An alternative way to evaluate $I_{mk}$
is to perform integration by parts \edt{in Eq.~(\ref{eq:Imk})}
\begin{equation}
    \label{Imk_alternative}
    I_{mk}(\E) = \frac{i k}{h \pi m} \int_{\phi_{\min}(\E)}^{\phi_{\max}(\E)}d\phi\; e^{i\frac{k}{h}\phi} \, \sin[m\psi(\E,\phi)],
\end{equation}
where the phase of synchrotron oscillations is given by Eq.~(\ref{eq:phase}).
For short bunches in a single rf system, $\phi(\E,\psi) \approx \sqrt{2\E}\cos\psi$ and the functions $I_{mk}(\E)$ can be approximated by the Bessel functions $J_m(x)$ of the first kind and the order $m$,
\begin{equation}
\label{eq:Imk_approx}
I_{m k} (\E) \approx i^m J_m\left(\frac{k}{h}\sqrt{2\E} \right).
\end{equation}
A comparison of the exact and approximate functions $I_{mk}$ is shown in Fig.~\ref{fig:Imk}.

\edt{The} perturbation of the line density at frequency $\Omega$ \edt{is}
\begin{equation}
    \label{eq:line_density_perturb}
    \Tilde{\lambda}(\phi,\Omega) = \sum_{k=-\infty}^{\infty} \Tilde{\lambda}_k (\Omega) e^{i \frac{k}{h} \phi},
\end{equation}
\edt{where} the harmonics $\Tilde{\lambda}_k(\Omega)$ are related to $\Tilde{\F}(\E,\psi,\Omega)$ as
\begin{equation}
    \label{eq:line_density_pert_harm}
    \Tilde{\lambda}_k (\Omega) = \frac{1}{2\pi h} \int_{-\pi h}^{\pi h}d\phi \, \Tilde{\lambda}(\phi,\Omega) e^{-i \frac{k}{h} \phi} =  \frac{\omega_{s0}^2}{2\pi h} \int_{-\pi}^{\pi} d\psi \int_0^{\E_{\max}} d\E \; \frac{\Tilde{\F}(\E,\psi,\Omega)}{\omega_{s}(\E)} \, e^{-i\frac{k}{h}\phi(\E,\psi)}.
\end{equation}
Here, the transformation of variables  $d\phi d \dot{\phi} = \omega_{s0}^2 \, d\psi \,d\E/\omega_s(\E)$ was used.

\edt{In} the last step, inserting Eq.~(\ref{eq:solution}) into Eq.~(\ref{eq:line_density_pert_harm}),  an infinite system of equations for \edt{the} harmonics of the line density perturbation can be obtained
\begin{equation}
    \label{eq:matrix_lebedev}
    \Tilde{\lambda}_p(\Omega) = \frac{\zeta}{h \cos\phi_{s0}} \sum_{k = -\infty}^{\infty} G_{p k} (\Omega) \, \frac{Z_k(\Omega)}{k} \, \Tilde{\lambda}_k(\Omega),
\end{equation}
where the beam transfer matrices~\edt{\cite{Shaposhnikova1994, Balbekov-Ivanov1986}} are defined as
\begin{equation}
    G_{p k} (\Omega) =  - i \, \omega_{s0}^2 \sum_{m=-\infty}^{\infty} \int_0^{\E_{\max}}\! d\E \frac{d\F(\E)}{d\E} \frac{\chk{I^*_{m p}(\E)} I_{m k}(\E)}{\Omega/m - \omega_s(\E)}. \label{eq:BTM}
\end{equation}
For real $\Omega$ inside the incoherent frequency band, $\Omega = m\omega_s(\E_m)$, the integrals in Eq.~(\ref{eq:BTM}) should be evaluated for $\mathrm{Im}\,\Omega \to -0$ (the boundary of the instability region) using the rule
\[
\lim_{\varepsilon \to +0} \frac{1}{x \chk{- i\varepsilon}} = \mathcal{P}\frac{1}{x} \chk{+ i\pi\delta(x)},
\]
where $\delta$ and $\mathcal{P}$ denote the Dirac delta function and the principal value of the integral, respectively.
Then, after substitution of a binomial distribution $\F$ from Eq.~(\ref{eq:binomial_distribution}), the element $G_{pk}$ becomes
\begin{align}
    G_{p k} (\Omega) &=  -\frac{\omega_{s0}}{2 A_N} \text{sgn}(\Omega) \sum_{m=1}^{\infty} \frac{ g^\prime(\E_m)}{\omega^\prime_s(\E_m)} I^*_{m p}(\E_m) I_{m k}(\E_m) \nonumber \\
    &- i \, \frac{\omega_{s0}}{\pi A_N} \sum_{m=1}^{\infty}  \mathcal{P}\int_0^{\E_{\max}} d\E g^\prime(\E) \frac{I^*_{m p}(\E) I_{m k}(\E) \omega_s(\E)}{\Omega^2/m^2 - \omega^2_s(\E)}, \label{eq:BTM_new}
\end{align}
where the first term exists only if the coherent frequency $\Omega$ belongs to the incoherent frequency band, i.e. $|\Omega| = m\omega_s(\E_m)$ for some $\E_m$ inside the bunch.
Note that Eq.~(\ref{eq:matrix_lebedev}), referred to below as \edt{the} Lebedev equation, is \edt{a} general equation\edt{,} since no approximations \edt{have been} used so far. In general, $G_{pk}$ depend on \edt{the} intensity parameter $\zeta$\edt{,} as they \edt{are} found after \edt{the} stationary problem is solved.
The solution for $\Omega$ exists if the determinant of the following matrix is zero
\begin{equation}
    \label{eq:determinant}
    D(\Omega,\zeta) = \det \left|\delta_{p k} - \frac{\zeta}{h\cos\phi_{s0}} \, G_{p k} (\Omega)\frac{Z_k(\Omega)}{k} \right| = \det\left|I - \zeta X(\Omega)\right| = 0,
\end{equation}
where $\delta_{pk}$ is the Kronecker delta, $I$ is the identity matrix and $X_{pk}(\Omega) = G_{pk}(\Omega) Z_k(\Omega)/(k h \cos\phi_{s0})$. The beam is unstable if the solution $\Omega$ has a negative imaginary part.
In the following sections this equation is applied to find the LLD threshold of a single bunch (Section~\ref{sec:Thesholds}) and the coupled-bunch instability thresholds and growth rates of a multi-bunch beam (Section~\ref{sec:multibunch}).

The exact solution of Eq.~(\ref{eq:determinant}) in a general case has to be obtained numerically, and the code MELODY (Matrix Equations for LOngitudinal beam DYnamics)~\cite{MELODY} was developed for this purpose~\cite{Ivan, Karpov2024}. \edt{The calculations follow the derivations above: first, for a given particle distribution function $\F(\E)$, bunch intensity, rf system and impedance model $Z(\omega)$, the stationary solution with the potential well distortion is found by the iterative procedure mentioned in Section~\ref{sec:main_equations}, together with the functions $\omega_s(\E)$, $\psi(\E,\phi)$ and $I_{mk}(\E)$; then the matrix elements $G_{pk}(\Omega)$ are evaluated from Eq.~(\ref{eq:BTM_new}) and the zeros of the determinant $D(\Omega,\zeta)$ are searched for, yielding the threshold intensity, at which a coherent mode emerges from the incoherent frequency band, as well as the frequencies, growth rates $-\mathrm{Im}\,\Omega$ and spectra $\Tilde{\lambda}_k$ of the coherent modes above the threshold.} In practice, the sum over $k$ and the sum over $m$ are truncated at some $k_{\max}$ and $m_{\max}$, sufficient for the convergence of the results.
The code treats a single bunch as well as a multi-bunch beam of equidistant bunches, each coupled-bunch mode $n$ being considered separately (Section~\ref{sec:multibunch}), in single and double rf systems. Contrary to the macro-particle simulations, which are not feasible for thousands of bunches\edt{,} the semi-analytic calculations are fast and allow the parameter space to be scanned for the existing and future accelerators.

\section{\label{sec:Thesholds} Loss of Landau damping}
\begin{figure}[htb]
	\centering
	\includegraphics[width = 0.55\textwidth]{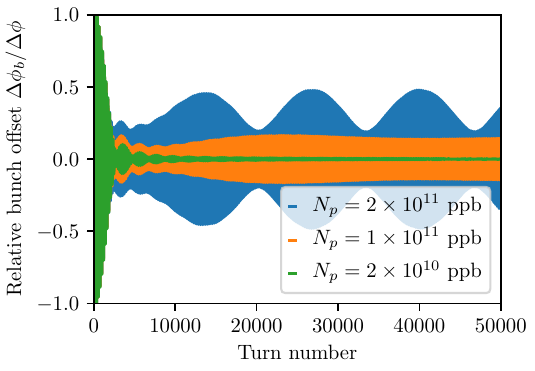}
	\caption{Offset evolution of the bunch with different intensities after a kick of $\Delta\phi=1$~deg\edt{, obtained from macro-particle simulations with the code BLonD~\cite{BLOND2020},} for a broad-band resonant impedance with $\mathrm{Im}Z/k = 0.07~\Omega$, $f_r = 4$~GHz and $Q=1$; the LHC parameters at injection energy, $V_0 = 6$~MV, zero-intensity $\tau_{4\sigma} = 0.82$~ns, and $\mu=2$. In this case the calculated LLD threshold $N_{p,th} \approx 5.7\times 10^{10}$. From Ref.~\cite{Ivan}.}
	\label{fig:LLD_kick}
\end{figure}

In the longitudinal plane, Landau damping is provided by the synchrotron frequency spread coming from \edt{the} nonlinearity of the rf field, modified by the self-induced wake field. At low intensities, the coherent bunch oscillations, excited, for example, by an injection error, are damped due to the decoherence of particles oscillating with different synchrotron frequencies. Landau damping is considered to be lost when the frequency of the coherent bunch oscillations moves outside the incoherent frequency band. In this case, the bunch oscillations are not damped anymore, which was observed in the Tevatron~\cite{Tevatron}, RHIC~\cite{RHIC}, SPS~\cite{SPS} and LHC~\cite{LHC}. Examples of the bunch offset evolution after a kick below and above the LLD threshold\edt{, obtained from macro-particle simulations with the code BLonD~\cite{BLOND2020},} are shown in Fig.~\ref{fig:LLD_kick}.

Analyzing solutions of the dispersion integral obtained from the Vlasov equation for infinite plasma, van Kampen~\cite{vKampen1, vKampen2} \edt{found} that they have a continuous and a discrete part. In our case, at low intensities ($\zeta \approx 0$), all van Kampen modes belong to the continuous spectrum, $\Omega = m\omega_s(\E)$, \edt{with singular eigenfunctions, and} Landau damping results from the phase mixing of these modes, which do not represent the collective motion of the particles. Above the threshold the discrete van Kampen modes\edt{, described by regular functions and lying outside $\omega_s(\E)$,} can emerge from the continuous spectrum, implying that Landau damping is lost~\cite{YHChin1983, burov2012van}.

\subsection{The LLD threshold for constant inductive impedance}
Here we derive an analytic expression for the LLD threshold using \edt{the} Lebedev equation~(\ref{eq:matrix_lebedev})\edt{, which had never been used in its original matrix form for this purpose until recently,} following \cite{Ivan}.
We will focus below on the case of $\eta \text{Im} Z/k >0$ (space charge below transition or inductance above transition), while the opposite sign is briefly discussed in \cite{Ivan}.

For the dipole mode ($m=1$), the LLD threshold is reached
when the frequency of the coherent dipole mode equals the maximum incoherent frequency $ \hat{\omega}_s = \max[\omega_s(\E)]$.
At low intensities, the~synchrotron frequency distribution in a single rf system is a monotonic function of the energy of synchrotron oscillations $\E$.
Assuming that \edt{this} still holds at the LLD threshold for a dipole mode $m=1$, we can search for the bunch intensity, at which $\Omega = \hat{\omega}_s = \omega_{s}(0)$ is a solution of Eq.~(\ref{eq:matrix_lebedev}).
\edt{Expanding the determinant (\ref{eq:determinant}) to the first order in the small parameter $\varepsilon \propto \zeta$, i.e. $\det(I - \zeta X) \approx 1 - \zeta\,\text{tr}(X)$, where $\text{tr}(X)$ is the~trace of the matrix $X$, one obtains} a general expression for the LLD threshold
\begin{equation}
    \label{eq:LLD_threshold_general}
    \zeta_\mathrm{th} \simeq \frac{1}{\text{tr}(X)} = h\cos\phi_{s0}\left[\sum_{k=-\infty}^{\infty} G_{k k} (\Omega)\frac{Z_k(\Omega)}{k}\right]^{-1}.
\end{equation}
Note that Eq.~(\ref{eq:LLD_threshold_general}) is exact if only one eigenvalue of the matrix $\zeta X$ is different from zero, and, in general, its accuracy is defined by the relative contribution of the other eigenvalues to the trace~\cite{Karpov2024}. \edt{The} potential well distortion can be neglected in \edt{the} calculation of $G_{kk}$\edt{,} as it would \edt{give} a higher-order term \edt{in} the~parameter $\varepsilon$.

\edt{The diagonal elements $G_{kk}$ can be evaluated analytically for short bunches ($\E_{\max} \ll 1$) in a~single rf system, using the approximation (\ref{eq:Imk_approx}) for $I_{mk}$ with only the dipole mode $m=1$ taken into account, $\omega_s(\E) = \omega_{s0}(1-\E/8)$ and the binomial distributions $g(\E) = (1-\E/\E_{\max})^\mu$ (for the details of the~calculation see Ref.~\cite{Ivan}). For a constant inductive impedance $Z_k = ik\,\text{Im}Z/k$, truncated at some harmonic $k_\edt{c}$, the sum in Eq.~(\ref{eq:LLD_threshold_general}) then yields the LLD threshold, positive for $\eta\,\text{Im}Z/k>0$, which can be expressed in terms of a generalized hypergeometric function of the parameter ${}y_{\max} = k_c \phi_{\max}/h$, where $\phi_{\max} = \sqrt{2\E_{\max}}$ is the maximum phase amplitude in the bunch~\cite{Ivan}. For ${}y_{\max} \gg 1$, a simple expression for the LLD threshold can be obtained:}
\begin{equation}
    \label{eq:threshold_binom_phimax_high_freq}
    \zeta_\mathrm{th} \approx -\frac{\pi \phi^4_{\max}h\cos\phi_{s0}}{32\mu(\mu+1)k_\edt{c}}\,\frac{1}{\text{Im}Z/k}\edt{,}
\end{equation}
\edt{so that the threshold is inversely proportional to the cutoff frequency $f_c = f_0 k_c$ and scales with the~fourth power of the bunch length. Without the truncation ($k_\edt{c} \to \infty$) the sum over $k$ diverges (except for $\mu = 0$): a constant inductive impedance above transition energy (the LHC case), or a capacitive one (space charge) below, leads to a zero LLD threshold. The finite thresholds obtained in the previous studies were due to the finite maximum frequency of the inductive impedance, naturally existing in numerical calculations; in reality, the reactive impedance $\text{Im}Z/k$ is always limited by the cutoff frequency of the~beam pipe.}

For $\mu=1/2$,
the Sacherer formalism \edt{yields} the LLD threshold \edt{$\zeta_1$} (in our notations)~\cite{NG} \edt{and} the~Hofmann-Pedersen approach~\cite{Hofmann1979}\edt{,} applied for short bunches\edt{,} yields \edt{$\zeta_2$:}
\[
\zeta_1 = -\frac{\phi^5_{\max}\cos\phi_{s0}}{18\,\text{Im}Z/k}\edt{, \hspace{1 cm}}
\zeta_2 = -\frac{\phi^5_{\max}\cos\phi_{s0}}{15\,\text{Im}Z/k}.
\]
\edt{These} expressions can be obtained from \edt{the general result} for ${}y_{\max} \approx 3.32 $ and ${}y_{\max} \approx 2.95 $, respectively.
This implies that to \edt{obtain} these  thresholds, the impedance \edt{in both cases} has to be truncated at the~frequency  $f_c \approx1/\tau_\mathrm{full}$, inversely proportional to the full bunch length $\tau_\mathrm{full} = 2\phi_{\max}/\omega_\mathrm{rf}$. \edt{They are therefore a low cutoff-frequency approximation ($f_c \lesssim 1/\tau_\mathrm{full}$, i.e. ${}y_{\max} \lesssim \pi$), in which the threshold scales with the fifth power of the bunch length; this approximation is not justified for} the constant inductive impedance $Z_k/k$\edt{, which} does not decay for $k_\edt{c}\xrightarrow{}\infty$.

\edt{The analytic threshold agrees well (within 30\% for $\mu > 1$) with the direct numerical solution of the Lebedev equation (\ref{eq:determinant}) obtained with the code MELODY~\cite{MELODY} for the LHC parameters at injection energy, as well as with the macro-particle simulations using the code BLonD~\cite{Ivan, BLOND2020}. The threshold curves for different $\mu$ are close to each other once the bunch length $\tau_{4\sigma}$, defined by Eq.~(\ref{eq:bunch_length}), is used, showing a~weak dependence on the details of the distribution.}

\edt{The results can also be applied to an arbitrary impedance $Z_k/k$ by introducing an effective impedance $(\text{Im}Z/k)_\mathrm{eff}$ with a corresponding effective cutoff harmonic $k_\mathrm{eff}$~\cite{Ivan, Karpov2024}; for a broad-band resonator impedance (\ref{eq:resonator}) with $Q=1$, one has $k_\mathrm{eff} = \omega_r/\omega_0$ and the dependence of the LLD threshold on the resonant frequency is the same as on the cutoff frequency of the truncated inductive impedance.}

\edt{Note also that there is a special class of the particle distribution functions with zero derivative in the bunch centre, $d\F/d\E = 0$ at $\E = 0$, which leads to a high and finite LLD threshold even for a~constant inductive impedance~\cite{Ivan}. In operation, such flat distributions are obtained as a result of the~specific rf manipulations (e.g. rf phase modulation), which were used to suppress bunch oscillations in the~Tevatron~\cite{CYTanABurov2012} and in the LHC~\cite{Shaposhnikova2014}.}

\subsection{Single-bunch instabilities}
The coherent modes, found as the solutions of the linearized Vlasov equation in the 2-D longitudinal phase space, are characterized by two mode numbers: the azimuthal mode number $m$ and the radial mode number, which is defined by the number of oscillations of the eigenfunction $\Tilde{\F}_m(\E,\Omega)$ within the bunch. Above the LLD threshold, the emerged coherent modes are undamped, but stable. The single-bunch instability driven by the broad-band impedance requires a significant potential well distortion, leading to the coupling of the modes with different or the same azimuthal mode numbers~\cite{Oide_Yokoya1990, Dyachkov, Karpov2023}. Its threshold is therefore typically higher than the LLD threshold. The main types of the single-bunch instability are
\begin{itemize}
    \item azimuthal mode coupling (the first mechanism proposed, in reality more rare);
    \item radial mode coupling;
    \item microwave instability (simultaneous excitation of many radial or azimuthal modes).
\end{itemize}
All of them were observed in the CERN SPS~\cite{Karpov2023}. Note that the potential well distortion is important for an accurate threshold calculation.
\newpage
\section{\label{sec:multibunch} Multi-bunch instabilities}
The multi-bunch instability is driven by a resonator impedance with a narrow bandwidth $\Delta\omega_r$, so that one bunch sees the wake field induced by the previous bunch(es). Bunch oscillations become coupled and this instability is also called the coupled-bunch instability (CBI). \edt{In this case,} the bunches of a train oscillate with the same frequency and amplitude, but with a phase of oscillation which changes from bunch to bunch by a constant value, so that the pattern of the bunch displacements looks like a wave propagating along the train. In the bunch profiles recorded turn after turn (mountain range display), the~instability is seen as growing oscillations of the bunch positions (dipole mode), which are correlated for the different bunches.

\subsection{Lebedev equation for a multi-bunch beam}
Let us consider a beam of $M$ identical and equidistant bunches, each containing $N_p$ particles. For such a~beam, the perturbation of the distribution function of the $j$-th bunch can be written as
\[
\Tilde{\F}_j(\E,\psi,t) = \Tilde{\F}(\E,\psi,\Omega)\, e^{i\Omega (t - jT_0/M)} e^{-i2\pi nj/M},
\]
where $n = 0, 1, ... M-1$ is the coupled-bunch mode number, which defines the phase shift $2\pi n/M$ between the oscillations of consecutive bunches.
Then the induced voltage is $M$ times larger than for a~single bunch and the perturbed line density contains only the harmonics $k = n + lM$, $-\infty < l < \infty$. As a result, the Lebedev equation~(\ref{eq:matrix_lebedev}) for the coupled-bunch mode $n$ becomes \edt{(see e.g. Appendix~A of Ref.~\cite{Karpov2024})}
\begin{equation}
    \label{eq:matrix_lebedev_MB}
    \Tilde{\lambda}_p(\Omega) = \frac{\zeta M}{h\cos\phi_{s0}} \sum_{k = n + lM} G_{p k} (\Omega) \, \frac{Z_k(\Omega)}{k} \, \Tilde{\lambda}_k(\Omega),
\end{equation}
where the beam transfer matrices $G_{pk}$ are defined by Eq.~(\ref{eq:BTM}) and $p$ also belongs to the same set of harmonics.

For a narrow-band resonator impedance (\ref{eq:resonator}) with the resonant frequency $\omega_r = k_r \omega_0$ and a bandwidth $\Delta\omega_r \ll M\omega_0$, only one term with the harmonic $k = n + lM$ closest to $k_r$ can be kept in Eq.~(\ref{eq:matrix_lebedev_MB}). Then the (infinite) system of equations reduces to a single equation and the dispersion relation for the~coherent frequency $\Omega$ is
\begin{equation}
    \label{eq:dispersion_NB}
    \frac{k}{Z_k(\Omega)} = \frac{\zeta M}{h\cos\phi_{s0}}\, G_{kk}(\Omega) = \xi\, G_{kk}(\Omega), \hspace{1 cm} \xi = \frac{2\pi h I_0}{V_0\cos\phi_{s0}},
\end{equation}
where $I_0 = M I_b = q N_p M f_0$ is the average beam current.

\subsection{Stability diagrams and instability threshold}
\begin{figure}[htb]
	\centering
	\includegraphics[width = 0.62\textwidth]{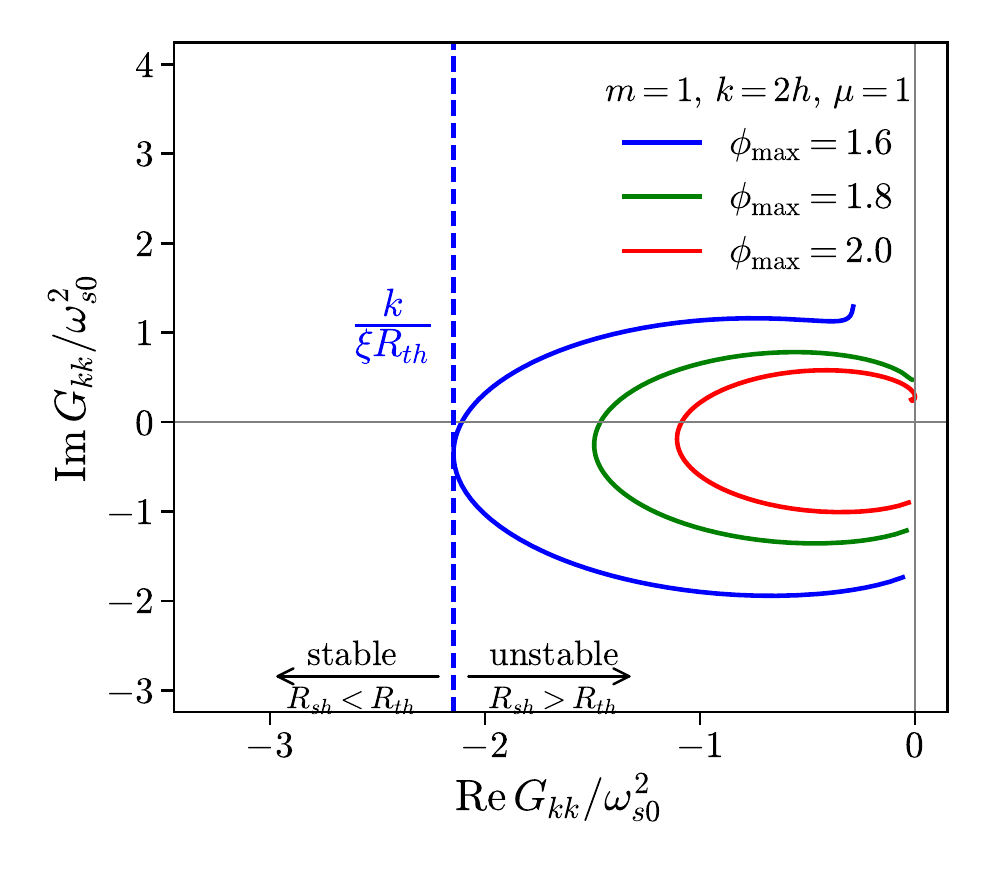}
	\caption{Examples of the stability diagram, the curve $G_{kk}(\Omega)/\omega_{s0}^2$ in the complex plane for real $\Omega$ inside the~incoherent frequency band, for the dipole mode ($m=1$) in a single rf system (stationary bucket) and $k = 2h$, calculated from Eqs.~(\ref{eq:ReG}) and (\ref{eq:ImG}) with the exact functions $I_{mk}(\E)$ and $\omega_s(\E)$ for the binomial distribution with $\mu = 1$ and three values of the bunch length, $\phi_{\max}= 1.6$, $1.8$ and $2.0$. Above transition energy $\xi <0$ and the~vertical line $\mathrm{Re}[k/(\xi Z_k)] = k/(\xi R_{sh})$ is in the left half-plane. The beam is stable if this line does not cross the~diagram, i.e. $|k/(\xi R_{sh})| > \max_\Omega |\mathrm{Re}\,G_{kk}(\Omega)|$. The dashed line, tangent to the diagram for $\phi_{\max} = 1.6$, corresponds to the~threshold shunt impedance $R_{th}$: for $R_{sh} < R_{th}$ the line $k/(\xi R_{sh})$ is to the left of it and the beam is stable, for $R_{sh} > R_{th}$ the line crosses the diagram and the beam is unstable.}
	\label{fig:stab_diagram}
\end{figure}

The stability boundary corresponds to $\text{Im}\,\Omega \to -0$ and the real part of the coherent frequency $\Omega$ inside the~incoherent frequency band, $\Omega = m\omega_s(\E_m)$. Then the real and imaginary parts of the diagonal elements $G_{kk}$ can be found from Eq.~(\ref{eq:BTM_new}),
\begin{align}
    \text{Re}\,G_{kk}(\Omega) &= -\frac{\omega_{s0}}{2 A_N} \text{sgn}(\Omega) \sum_{m=1}^{\infty} \frac{ g^\prime(\E_m)}{\omega^\prime_s(\E_m)} \left|I_{m k}(\E_m)\right|^2, \label{eq:ReG}\\
    \text{Im}\,G_{kk}(\Omega) &= -\frac{\omega_{s0}}{\pi A_N} \sum_{m=1}^{\infty}  \mathcal{P}\int_0^{\E_{\max}} d\E g^\prime(\E) \frac{\left|I_{m k}(\E)\right|^2 \omega_s(\E)}{\Omega^2/m^2 - \omega^2_s(\E)}. \label{eq:ImG}
\end{align}
The stability diagram\edt{, first used in Ref.~\cite{Balbekov-Ivanov1986} for a different particle distribution function,} is the contour in the complex plane obtained from the right-hand side of Eq.~(\ref{eq:dispersion_NB}) by varying $\Omega$ inside the incoherent frequency band, see an example in Fig.~\ref{fig:stab_diagram}.

For a resonator impedance the left-hand side is
\[
\frac{\edt{1}}{Z_k(\Omega)} = \frac{\edt{1}}{R_{sh}}\left[1 + iQ\left(\frac{\omega}{\omega_r} - \frac{\omega_r}{\omega}\right)\right], \hspace{1 cm} \omega = k\omega_0 + \Omega,
\]
and the threshold is defined by the crossing of the stability diagram with the vertical line at $\text{Re}[k/Z_k] = k/R_{sh}$.
Therefore, the following conditions are satisfied simultaneously at the threshold:
\begin{align}
    \frac{k}{R_{sh}} &= \xi\, \text{Re}\,G_{kk}(\Omega),\\
    \frac{k}{R_{sh}}\, Q\left(\frac{\omega}{\omega_r} - \frac{\omega_r}{\omega}\right) &= \xi\, \text{Im}\,G_{kk}(\Omega).
\end{align}
If the resonator bandwidth $\Delta\omega_r \gg \omega_{s0}$, the left-hand side of the second equation weakly depends on $\Omega$ and it is almost zero. In this case, the lowest instability threshold corresponds to the maximum value of $\xi\text{Re}\,G_{kk}(\Omega)$, so that the beam is stable if
\begin{equation}
    \label{eq:Rsh_threshold}
    R_{sh} < \frac{k_r}{\max_\Omega\left[\xi\, \text{Re}\,G_{k_r k_r}(\Omega)\right]} = \frac{V_0 |\cos\phi_{s0}| }{2\pi^2 h I_0 \omega_{s0}^2}\, k_r\, \chk{\min_\Omega \left[\sum_{m=1}^{\infty} \frac{\F^\prime(\E_m) |I_{m k_r}(\E_m)|^2}{\omega^\prime_s(\E_m)}\right]^{-1}},
\end{equation}
where \edt{$\E_m$ is the energy of synchrotron oscillations of the particles in resonance with the coherent mode, $|\Omega| = m\omega_s(\E_m)$, and,} according to Eqs.~(\ref{eq:binomial_distribution}), $g^\prime = 2\pi\omega_{s0}A_N \F^\prime$ \edt{was used}. This expression defines the~requirements for the damping of the HOMs in the rf cavities for a given beam current $I_0$.

For a stationary bucket (no acceleration) above transition energy ($\phi_{s0} = \pi$), a binomial distribution $g(\E) = (1 - \E/\E_{\max})^\mu$ and short bunches, neglecting the potential well distortion and using the~approximation (\ref{eq:Imk_approx}) with $\omega_s(\E) = \omega_{s0}(1 - \E/8)$, Eq.~(\ref{eq:ReG}) becomes
\begin{equation}
    \label{eq:ReG_binom}
    \text{Re}\,G_{kk}(\Omega) = - \frac{16\mu(\mu+1)}{\phi_{\max}^4}\, \text{sgn}(\Omega) \sum_{m=1}^{\infty} \left(1 - \frac{\E_m}{\E_{\max}}\right)^{\mu -1} J_m^2\left(\frac{k}{h}\sqrt{2\E_m}\right).
\end{equation}
The analysis of this expression shows~\cite{Karpov2024} that for $\mu \ge 1$ and any $k_r$ the instability threshold is always the~lowest for the dipole mode $m=1$, and the contribution of higher-order azimuthal modes is negligible, since the full synchrotron frequency spread, equal to $\omega_{s0}\E_{\max}/8$, is smaller than $\omega_{s0}/2$. Then, in terms of the full bunch length $\tau$, with $\phi_{\max} = \pi f_\mathrm{rf}\tau$, and the resonant frequency $f_r = k_r f_0$, the threshold (\ref{eq:Rsh_threshold}) can be written in the form~\cite{Karpov2019}
\begin{equation}
    \label{eq:Rsh_threshold_binom}
    R_{sh} < \frac{V_0 \left(\pi f_\mathrm{rf}\tau\right)^3}{32\, I_0}\, W_\mu(f_r\tau), \hspace{1 cm}
    W_\mu(x) = \frac{x}{\mu(\mu+1)} \min_{\edt{0\le\E_m\le\E_{\max}}} \frac{\left(1 - \edt{\E_m/\E_{\max}}\right)^{1-\mu}}{\chk{J_1^2(\pi x \edt{\sqrt{\E_m/\E_{\max}}})}}.
\end{equation}
For a flat distribution with zero derivative in the bunch centre, the threshold was found in Ref.~\cite{Balbekov-Ivanov1986}. In this case, above a certain value of the resonant frequency, the higher-order azimuthal modes ($m > 1$) have the lowest instability threshold.

\begin{figure}[htb]
	\centering
	\includegraphics[width = 0.5\textwidth]{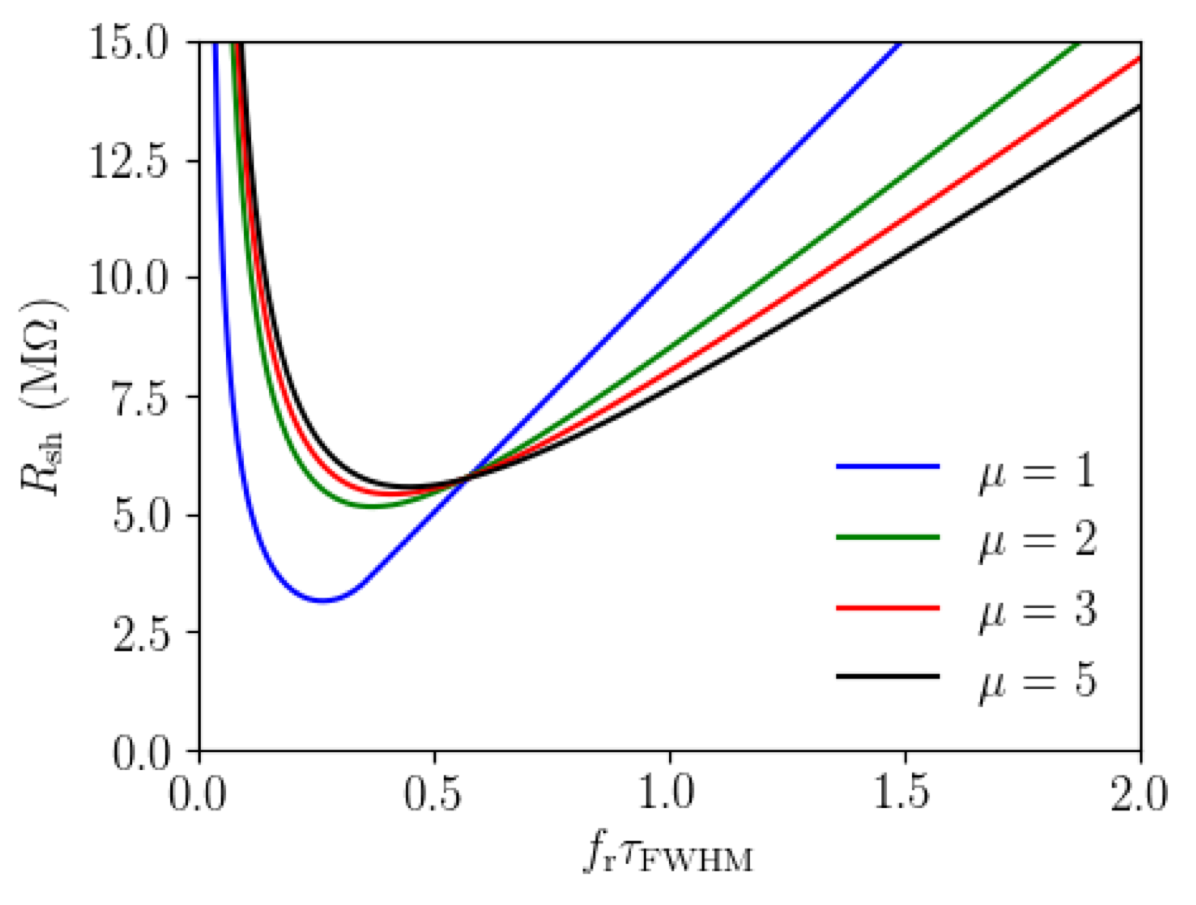}
	\caption{Threshold shunt impedance of the coupled-bunch instability, Eq.~(\ref{eq:Rsh_threshold_binom}), as a function of the resonant frequency multiplied by the FWHM bunch length for the binomial distributions with different $\mu$. Example for FCC-hh at 50~TeV with $N_p = 10^{11}$, $V_0 = 38$~MV and $\gamma_\mathrm{tr} = 99.3$. From Ref.~\cite{Karpov2019}.}
	\label{fig:Rsh_fcc}
\end{figure}

An example of the threshold shunt impedance calculated from Eq.~(\ref{eq:Rsh_threshold_binom}) for the FCC-hh parameters is shown in Fig.~\ref{fig:Rsh_fcc} as a function of $f_r\tau_\mathrm{FWHM}$ for different values of $\mu$\edt{~\cite{Karpov2019}}.

Note that, in practice, the full bunch length is difficult to determine due to the noise in the measured signals and the bunch length is usually defined by the FWHM. For the same FWHM bunch length, the~thresholds are very similar for $\mu >1$, with the minimum in the region $f_r \tau_\mathrm{FWHM} \sim 0.5$, while for $\mu = 1$ (parabolic bunch) the minimum threshold is about 50\% lower. The threshold shunt impedance grows almost linearly with the resonant frequency for $f_r\tau_\mathrm{FWHM} > 1$.

\subsection{Impact of the broad-band impedance on the coupled-bunch instability threshold}
So far, the thresholds of the coupled-bunch instability and of the loss of Landau damping were calculated separately. In reality, both narrow-band and broad-band impedances are always present in the ring and the~general equation (\ref{eq:matrix_lebedev_MB}) allows the beam stability to be analyzed self-consistently~\cite{Karpov2024}.
Following the~same approach as in Section~\ref{sec:Thesholds}, i.e. keeping only the first-order term in the expansion of the determinant, the threshold intensity parameter for the coherent mode $\Omega$ can be written as
\begin{equation}
    \label{eq:threshold_trace_MB}
    \frac{1}{\zeta_\mathrm{th}(\Omega)} \simeq \text{tr}(X) = \frac{M}{h\cos\phi_{s0}} \sum_{k = n+lM} G_{kk}(\Omega)\frac{Z_k(\Omega)}{k} = \frac{1}{\zeta^{NB}(\Omega)} + \frac{1}{\zeta^{BB}(\Omega)},
\end{equation}
where the sum over $k$ was split into the term $k = k_r$ with the narrow-band impedance and all other terms with the broad-band impedance,
\begin{equation}
    \frac{1}{\zeta^{NB}(\Omega)} = \frac{M}{h\cos\phi_{s0}}\, G_{k_r k_r}(\Omega)\frac{Z_{k_r}(\Omega)}{k_r}, \hspace{1 cm}
    \frac{1}{\zeta^{BB}(\Omega)} = \frac{iM}{h\cos\phi_{s0}} \sum_{k \ne k_r} G_{kk}(\Omega)\frac{\text{Im}Z_k}{k}.
\end{equation}
Here, the contribution of the real part of the broad-band impedance in the sum is close to zero due to the symmetry $G_{kk} = G_{-k-k}$.
For the narrow-band impedance only, expression (\ref{eq:threshold_trace_MB}) is exact, since the matrix has only one non-zero element, and it gives the coupled-bunch instability threshold $\zeta^{NB}_\mathrm{th} = \zeta^{NB}(\Omega_{NB})$ discussed in the previous subsection. For the broad-band impedance only, it gives the LLD threshold $\zeta^{BB}_\mathrm{th} = \zeta^{BB}(\Omega_{BB})$ with $\Omega_{BB} = \omega_s(0)$, as derived in Section~\ref{sec:Thesholds}, with a maximum error of a~factor of two.
The coherent mode $\Omega_g$ for the general case differs from $\Omega_{NB}$ and $\Omega_{BB}$, found for each impedance separately, and Eq.~(\ref{eq:threshold_trace_MB}) should be solved numerically. If $\zeta^{BB}_\mathrm{th} \ll \zeta^{NB}_\mathrm{th}$, then $\Omega_g \approx \Omega_{BB}$, and the instability threshold is reduced with respect to $\zeta^{NB}_\mathrm{th}$ and can \edt{even be} below the LLD threshold~\cite{Karpov2024}. In the opposite case, the broad-band impedance has a negligible impact and $\Omega_g \approx \Omega_{NB}$.
Nevertheless, an~approximate threshold
\begin{equation}
    \label{eq:generalized_threshold}
    \frac{1}{\zeta_\mathrm{th}} \approx \frac{1}{\zeta^{NB}_\mathrm{th}} + \frac{1}{\zeta^{BB}_\mathrm{th}}
\end{equation}
describes well the results of the full self-consistent calculations~\cite{Karpov2024}, with the error not exceeding the~one for the LLD threshold. This generalized expression shows the key role of the LLD threshold and demonstrates how two impedance contributions add up.

The resulting multi-bunch instability has a mechanism different from the CBI driven by the~narrow-band impedance only. This can be illustrated by the character of the perturbation in the~longitudinal phase space~\cite{Karpov2024}. For the narrow-band impedance only, the perturbation mainly involves the high-amplitude particles, while the LLD mode due to the broad-band impedance is located in the~bunch centre. For the new mechanism of the coupled-bunch instability in the presence of the broad-band impedance, the mode is also localized in the bunch centre.

\begin{figure}[htb]
	\centering
	\includegraphics[width = 0.55\textwidth]{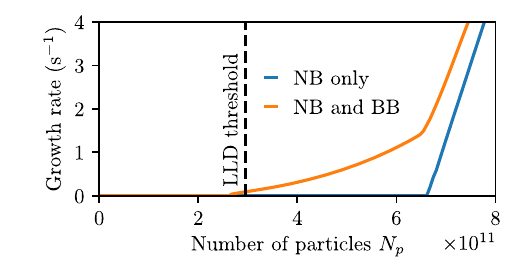}
	\caption{Growth rates of the multi-bunch instability versus bunch intensity calculated for the LHC at 450~GeV with $M = h/10$, $\phi_{\max} = 2.4$, and the NB impedance with $R_{sh} = 280$~k$\Omega$, $f_r = 582$~MHz, and $Q = 1360$ in two cases: with (orange) and without (blue) the BB impedance having $R_{sh} = 38$~k$\Omega$, $f_r = 5$~GHz, and $Q=1$. The~LLD threshold is shown by the dashed line. Calculations with the code MELODY\edt{~\cite{MELODY}}, from Ref.~\cite{Karpov2024}.}
	\label{fig:cbi_examples}
\end{figure}

These findings, confirmed by macro-particle simulations with the code BLonD\edt{~\cite{BLOND2020}}, are important for the existing and future accelerators. For example, in the LHC the lowest CBI threshold due to a HOM of the future crab cavities ($R_{sh} = 280$~k$\Omega$, $f_r = 582$~MHz, $Q = 1360$) calculated in the presence of the~broad-band impedance is three times lower than for the HOM only, see Fig.~\ref{fig:cbi_examples}.

In the SPS, the CBI is driven by a HOM of the main 200~MHz rf system at $f_r \approx 915$~MHz. For the~fixed-target beams, which fill the whole SPS ring ($M = h$), the CBI thresholds found for the realistic SPS impedance model and for the HOM impedance only practically coincide~\cite{Karpov2024}, since in this case the~CBI threshold is lower than the LLD threshold by an order of magnitude due to a low single-bunch intensity. If the bunches fill every fifth rf bucket (as LHC-type beams), the HOM-driven CBI threshold is higher and the resulting threshold is more affected by the broad-band part of the SPS impedance. This explains why the LHC bunch train in the SPS has a very low CBI threshold, which also weakly depends on the number of bunches in a train. To deliver high-intensity LHC beams, an additional 800~MHz rf system, increasing the synchrotron frequency spread and the CBI threshold, is routinely deployed in operation, see Section~\ref{sec:cures}.

\subsection{Growth rates}
\begin{figure}[htb]
	\centering
	\includegraphics[width = 0.55\textwidth]{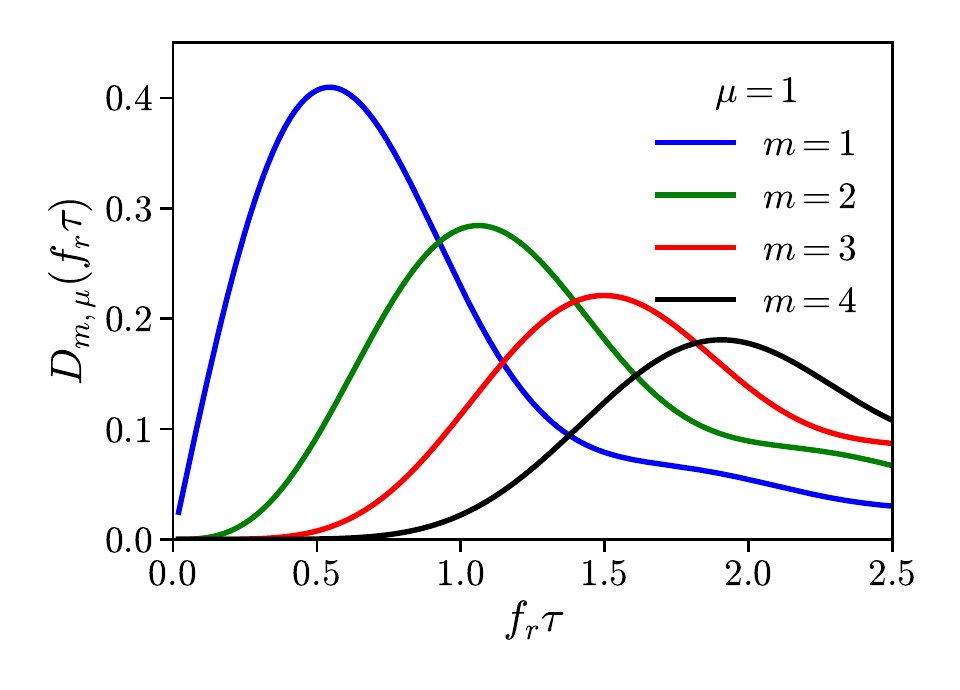}
	\caption{Form factor $D_{m,\mu}(f_r\tau)$ in the growth rate, Eq.~(\ref{eq:growth_rate}), for $\mu = 1$ and the azimuthal modes $m = 1, 2, 3, 4$ as a function of the resonant frequency multiplied by the full bunch length (see also Ref.~\cite{Shaposhnikova1999}).}
	\label{fig:formfactor}
\end{figure}

The instability growth rate $1/\tau_\mathrm{inst} = -\text{Im}\,\Omega$ for a separate azimuthal mode $m$ (no mode coupling) can be easily found \edt{by} neglecting the synchrotron frequency spread $\Delta\omega_s$, which is justified for $|\text{Im}\,\Omega| \gg \Delta\omega_s$. In this case, keeping only the resonant term in Eq.~(\ref{eq:BTM}), we have
\begin{equation}
    G_{kk}(\Omega) \simeq -i\omega_{s0}^2\, \frac{m}{\Omega - m\omega_{s0}} \int_0^{\E_{\max}} d\E\, \F^\prime(\E) |I_{mk}(\E)|^2,
\end{equation}
so that the dispersion relation (\ref{eq:dispersion_NB}) gives the coherent frequency shift
\begin{equation}
    \label{eq:growth_general}
    \Omega - m\omega_{s0} = -i\xi\omega_{s0}^2\, m\, \frac{Z_k(\Omega)}{k} \int_0^{\E_{\max}} d\E\, \F^\prime(\E) |I_{mk}(\E)|^2,
\end{equation}
where $\xi$ is defined in Eq.~(\ref{eq:dispersion_NB}). Since $\F^\prime(\E) <0$, the imaginary part of the coherent frequency shift is $\text{Im}\,\Omega \propto \xi\, \text{Re}Z_k/k$, i.e. above transition energy ($\xi<0$) the mode is unstable for the harmonic $k>0$, which corresponds to the upper synchrotron sideband $k\omega_0 + m\omega_s$ of the resonant frequency, while below transition ($\xi>0$) the unstable mode contains the harmonic $k<0$ (the lower sideband).
For the binomial distribution (\ref{eq:binomial_distribution}) with the normalization factor $A_N \simeq \E_{\max}/(\mu+1)$ and short bunches, after substitution of $\F^\prime(\E)$, $I_{mk}$ from Eq.~(\ref{eq:Imk_approx}), $\E_{\max} = \phi_{\max}^2/2$ and $\phi_{\max} = h\omega_0\tau/2$, we finally obtain for the unstable mode
\begin{equation}
    \label{eq:growth_rate}
    \frac{1}{\tau_\mathrm{inst}\,\omega_{s0}} = \chk{-\frac{\text{Im}\,\Omega}{\omega_{s0}}} = \frac{4}{\pi^2}\, \frac{I_0\, \text{Re}Z_{k}}{h V_0 |\cos\phi_{s0}|}\, \frac{D_{m,\mu}(f_r\tau)}{f_0\tau},
\end{equation}
with the form factor
\begin{equation}
    \label{eq:formfactor}
    D_{m,\mu}(f_r\tau) = \frac{m\mu(\mu+1)}{f_r\tau} \int_0^1 x \left(1 - x^2\right)^{\chk{\mu-1}} J_m^2(\pi f_r \tau x)\, dx,
\end{equation}
shown in Fig.~\ref{fig:formfactor} for $\mu=1$.

The growth rate is proportional to the beam current and the resistive impedance at the frequency of the coupled-bunch mode, $k\omega_0 + m\omega_{s0}$, and it is inversely proportional to the rf voltage and the bunch length. For a given bunch length $\tau$, the maximum growth rate of the azimuthal mode $m$ is at $f_r\tau \approx m/2$, so that for $f_r \tau < 1$ mainly dipole oscillations are excited, while the higher multipoles become dominant for $f_r \tau > 1$.

Note that the harmonics $k_r = n + lM$ and $-k_r = (M - n) + (-l-1)M$, which correspond to the upper and lower synchrotron sidebands of the resonant frequency, $\omega_r \pm m\omega_s$, belong to the different coupled-bunch modes, $n$ and $M-n$ (for $n \ne 0$ and $n \ne M/2$). Since $Z_{-k}(\Omega)/(-k) = -Z^*_{k}(-\Omega)/k$, one of these two modes is unstable and the other one is damped, depending on the sign of $\eta$: above transition energy the unstable mode is the one with the line $k_r>0$ in its spectrum, and below transition it is the mode $M-n$ with the line $-k_r$.

\subsection{Robinson instability}
The situation is different for the impedance at the fundamental rf frequency $\omega_r \simeq h\omega_0$, e.g. the impedance of the accelerating cavities, seen by all bunches, or, more generally, for $k_r$ close to a multiple of $M/2$. In this case the two harmonics $k_1 = n + l_1M$ and $k_2 = n + l_2 M$, where $l_2 = -l_1$ or $l_2 = -l_1 - 1$, so that $k_2 = -k_1$ for $n = 0$ or $n = M/2$, are equally close to the resonant frequency and both terms should be kept in the Lebedev equation~(\ref{eq:matrix_lebedev_MB}). For the fundamental rf impedance, $k_1 = -k_2 = h$ and this is the coupled-bunch mode $n=0$ (all bunches oscillate in phase), which is the only mode existing also for a~single bunch.
Then, instead of Eq.~(\ref{eq:growth_general}), we have the system of two equations
\begin{align}
    (\Omega - m\omega_{s0})\Tilde{\lambda}_{k_1} &= -i\xi\omega_{s0}^2\, m \left(g_{11}\frac{Z_{k_1}}{k_1}\Tilde{\lambda}_{k_1} + g_{12}\frac{Z_{k_2}}{k_2}\Tilde{\lambda}_{k_2}\right), \nonumber\\
    (\Omega - m\omega_{s0})\Tilde{\lambda}_{k_2} &= -i\xi\omega_{s0}^2\, m \left(g_{21}\frac{Z_{k_1}}{k_1}\Tilde{\lambda}_{k_1} + g_{22}\frac{Z_{k_2}}{k_2}\Tilde{\lambda}_{k_2}\right),
    \label{eq:robinson_system}
\end{align}
where
\[
g_{12} \equiv g^m_{k_1k_2} = \int_0^{\E_{\max}} d\E\, \F^\prime(\E)\, I_{mk_1}(\E) I^*_{mk_2}(\E).
\]
Taking into account that $g_{11} \simeq g_{22}$ and $g_{12} \simeq g_{21} = (-1)^m g_{11}$ (for a symmetric potential well, see the~properties of $I_{mk}$ in Section~\ref{sec:perturbation}), the determinant of the system (\ref{eq:robinson_system}) gives two solutions. The first one, $\Omega = m\omega_{s0}$, corresponds to the mode which is not affected by the impedance (the two sidebands cancel each other), and the second one is
\begin{equation}
    \Omega - m\omega_{s0} \simeq -i\xi\omega_{s0}^2\, m\, g_{11} \left(\frac{Z_{k_1}}{k_1} + \frac{Z_{k_2}}{k_2}\right) = -i\xi\omega_{s0}^2\, m\, g_{11}\, \frac{Z(k_1\omega_0 + \Omega) - Z^*(k_1\omega_0 - \Omega)}{k_1},
    \label{eq:robinson_solution}
\end{equation}
where we used that $Z_{k_2}/k_2 = Z(-k_1\omega_0 + \Omega)/(-k_1) = -Z^*(k_1\omega_0 - \Omega)/k_1$ for real $\Omega$. Since $g_{11} < 0$, the~imaginary part of the coherent frequency shift is $\text{Im}\,\Omega \propto \xi\,[\text{Re}Z(k_1\omega_0 + m\omega_s) - \text{Re}Z(k_1\omega_0 - m\omega_s)]$.
Thus, contrary to the case of the narrow-band impedance considered in the previous subsection, the~growth rate is now defined not by the impedance at one sideband, but by the difference of the resistive impedance at the upper and lower synchrotron sidebands of the rf frequency. There is no instability if the impedance is the same at the two sidebands, and this is why the dependence $Z_k(\Omega) = Z(k\omega_0 + \Omega)$ of the impedance on the coherent frequency, neglected so far, should be taken into account here. This is the Robinson instability~\cite{Robinson1964}, which is one of the first instabilities of bunched beams described theoretically. Using the same short-bunch approximation as in the derivation of Eq.~(\ref{eq:growth_rate}), its growth rate for the azimuthal mode $m$ can be written as
\begin{equation}
    \label{eq:robinson_rate}
    \frac{1}{\tau_\mathrm{inst}\,\omega_{s0}} = \chk{-\frac{\text{Im}\,\Omega}{\omega_{s0}}} = \chk{\text{sgn}(\eta)}\, \frac{4}{\pi^2}\, \frac{I_0\, \left[\text{Re}Z(h\omega_0 + m\omega_{s}) - \text{Re}Z(h\omega_0 - m\omega_{s}) \right]}{h V_0 |\cos\phi_{s0}|}\, \frac{D_{m,\mu}(f_\mathrm{rf}\tau)}{f_0\tau},
\end{equation}
with the form factor $D_{m,\mu}$ from Eq.~(\ref{eq:formfactor}), where we used that $\text{sgn}(\xi) = -\text{sgn}(\eta)$. The sign of the growth rate depends on the sign of the slip factor $\eta$: above transition energy the beam is unstable if the resistive impedance at the upper sideband is larger than at the lower one, and the opposite is true below transition. The physical reason is that above transition the particles with higher energy have a lower revolution frequency and the energy exchange between the beam and the impedance at the two sidebands changes the sign.

For the cavity impedance in the form of the resonator (\ref{eq:resonator}) with $Q \gg 1$ and a small detuning $\Delta\omega = \omega_r - h\omega_0$, $|\Delta\omega| \ll \omega_r$, the difference of the resistive impedances at the sidebands of the dipole mode can be expressed as
\begin{equation}
    \text{Re}Z(h\omega_0 + \omega_{s}) - \text{Re}Z(h\omega_0 - \omega_{s}) \simeq 2\omega_s \left.\frac{d\,\text{Re}Z}{d\omega}\right|_{h\omega_0} = \frac{16 R_{sh} Q^2 \omega_s\, \Delta\omega}{\omega_r^2 \left(1 + 4Q^2\Delta\omega^2/\omega_r^2\right)^2},
\end{equation}
where we assumed $\omega_s \ll \Delta\omega_r$. Then, according to Eq.~(\ref{eq:robinson_rate}), the growth rate is proportional to $\text{sgn}(\eta)\,\Delta\omega$ and the well-known Robinson stability criterion follows: the cavity should be detuned below the rf frequency, $\omega_r < h\omega_0$, above transition energy and above the rf frequency, $\omega_r > h\omega_0$, below transition. For the correct sign of the detuning the cavity impedance provides the so-called Robinson damping of the~coherent dipole oscillations, which is used, for example, to damp the injection errors. Note that the~cavity detuning is anyway required in the presence of the beam loading in order to minimize the rf power (see \edt{also the} lecture on beam loading by H.~Damerau at this School), and it has the sign required by the Robinson criterion. At high beam loading, when the beam-induced voltage in the cavity becomes comparable with the rf voltage, the coherent frequency shift (\ref{eq:robinson_solution}) is not small anymore, the real part of the coherent frequency is also modified, and the second (high-intensity) Robinson criterion limits the beam current for a given detuning. In this regime the beam stability is also strongly affected by the rf feedback and tuning loops~\cite{Pedersen1975}.

\subsection{Spectrum of the unstable beam}
\begin{figure}[htb]
	\centering
	\includegraphics[width = 0.85\textwidth]{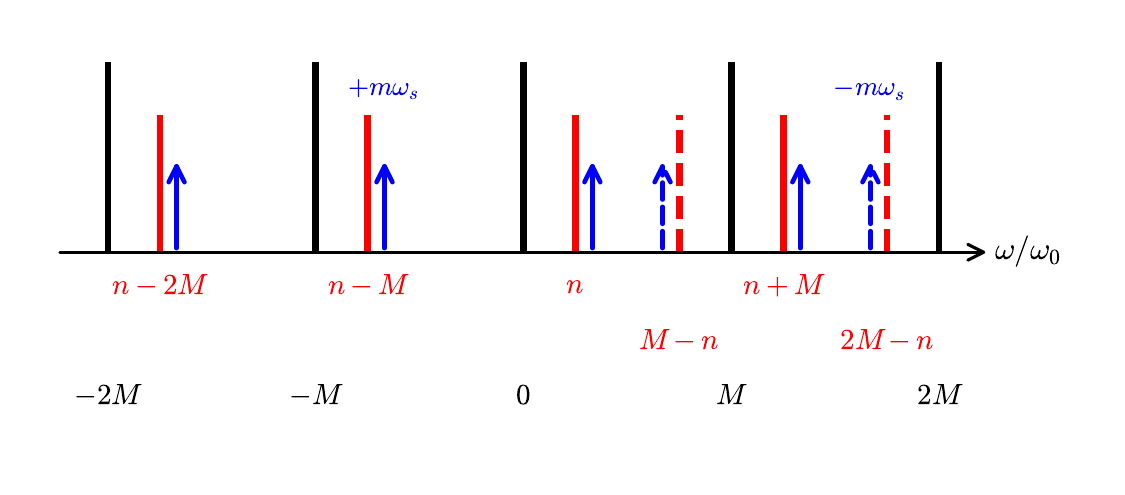}
	\caption{Spectrum of the coupled-bunch mode $n$ for $M$ equidistant bunches as it is seen on a spectrum analyzer: the~beam lines at multiples of $M\omega_0$ (black), the lines at $(n + lM)\omega_0$ (red, solid) with the upper synchrotron sidebands $+m\omega_s$ (blue arrows) and the lines at $[(l+1)M - n]\omega_0$ (red, dashed), corresponding to the negative frequencies, with the lower sidebands $-m\omega_s$ (dashed arrows).}
	\label{fig:spectrum_sketch}
\end{figure}

For a narrow-band resonant impedance at \edt{an} unknown resonant frequency $\omega_r = p_r\omega_0$, the spectrum of the unstable beam has components at the frequencies
\begin{equation}
    \omega = (n + lM)\omega_0 + m\omega_s, \hspace{1 cm} -\infty < l < \infty,
\end{equation}
where $n = 0, 1, ... M-1$ is the coupled-bunch mode number, $M$ is the number of equidistant bunches in the ring and $m = 1, 2, ...$ is the multipole (azimuthal) number. On the spectrum analyzer, negative frequencies appear at $[(l+1)M - n]\omega_0 - m\omega_s$, so that for each beam line $lM\omega_0$ one can see two lines at $(n + lM)\omega_0$ and $[(l+1)M - n]\omega_0$ with the synchrotron sidebands, see Fig.~\ref{fig:spectrum_sketch}.

\begin{figure}[htb]
	\centering
	\begin{tabular}{cc}
	\includegraphics[width = 0.46\textwidth]{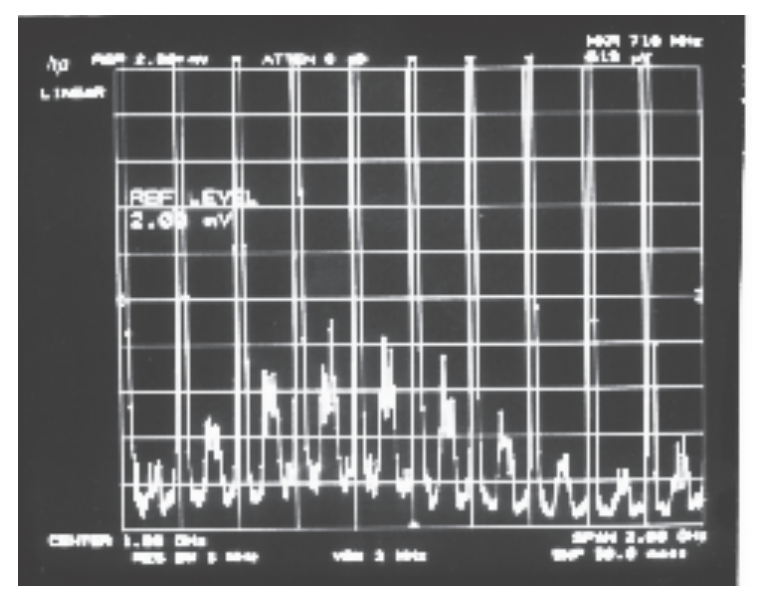} &
	\includegraphics[width = 0.42\textwidth]{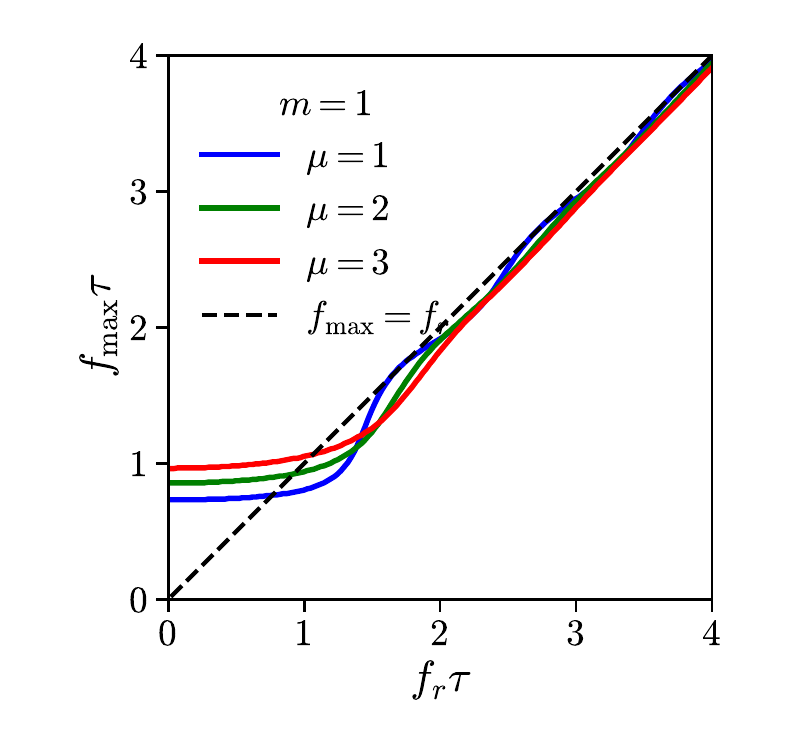}
	\end{tabular}
	\caption{Left: measured spectrum (0--2~GHz) of the unstable beam with 5~ns bunch spacing in the CERN SPS~\cite{Shaposhnikova1999}. Right: the frequency $f_{\max}$ of the maximum of the spectrum envelope multiplied by the full bunch length $\tau$ as a~function of $f_r\tau$ for the dipole mode $m=1$ driven by a narrow-band impedance, together with the line $f_{\max} = f_r$ (dotted), after Ref.~\cite{Shaposhnikova1999}. The envelope $|\Tilde{\lambda}_p| \propto |G_{pk_r}|$ was calculated from Eq.~(\ref{eq:BTM}) in the short-bunch approximation, neglecting the synchrotron frequency spread, for the binomial distributions with $\mu = 1, 2, 3$. For $f_r\tau \lesssim 1$ the~spectrum maximum is defined by the bunch length only ($f_{\max}\tau \simeq 0.8$--$1$, slightly depending on the bunch shape), while for $f_r\tau \gtrsim 2$ it follows the resonant frequency.}
	\label{fig:SPS_spectrum}
\end{figure}

Therefore, the measured mode number $n$ is not sufficient to determine $\omega_r$, since $n + lM \approx \pm p_r$ and $l$ is unknown. To identify the resonant frequency of the impedance driving the instability, one can measure the mode number $n$ for different $M$ (different bunch spacing) or measure the frequency $f_{\max}$ of the maximum of the spectrum envelope~\cite{Shaposhnikova1999}. As can be seen in Fig.~\ref{fig:SPS_spectrum} (right), if $f_{\max}\tau < 1$, the resonant frequency $f_r < 1/\tau$, and if $f_{\max}\tau > 1$, then $f_r \simeq f_{\max}$.

An example of the measured spectrum of the unstable beam with 5~ns bunch spacing in the CERN SPS, with the beam lines at multiples of 200~MHz, is shown in Fig.~\ref{fig:SPS_spectrum} (left). In the SPS, the resonant \edt{frequency} of the \edt{HOM} of the 200~MHz travelling wave cavities at 912~MHz \edt{was} identified from the spectrum measurements\edt{. The} shunt impedances can be estimated from the measurements of the instability growth rates using Eq.~(\ref{eq:growth_rate}).

\section{\label{sec:cures} Cures of longitudinal beam instabilities}
The possible cures of the longitudinal beam instabilities can be \edt{conditionally} divided into two groups: active and passive.
The active cures include
\begin{itemize}
    \item feedback systems;
    \item higher harmonic rf system;
    \item controlled longitudinal emittance blow-up.
\end{itemize}
The passive cures are
\begin{itemize}
    \item HOM damping (couplers);
    \item HOM-free cavity design;
    \item impedance reduction (modification of machine elements);
    \item change of optics (of $\gamma_\mathrm{tr}$);
    \item synchrotron radiation damping in lepton rings.
\end{itemize}
\edt{This division is not absolute: for example, the higher harmonic rf system can be used in a passive mode.}
The feedback systems are discussed in dedicated lectures at this School. The last two active cures, namely, the higher harmonic rf system and the controlled emittance blow-up (during the acceleration cycle or, for the LHC, also in physics), rely on the increase of the synchrotron frequency spread and therefore of Landau damping. They are widely used in the CERN accelerators.
\edt{Note that in the SPS the~instability threshold decreases with beam energy for a constant longitudinal emittance, which explains the requirement for the longitudinal emittance blow-up along the acceleration cycle with the rf voltage limited by the available rf power.}

\begin{figure}[htb]
	\centering
	\includegraphics[width = 0.95\textwidth]{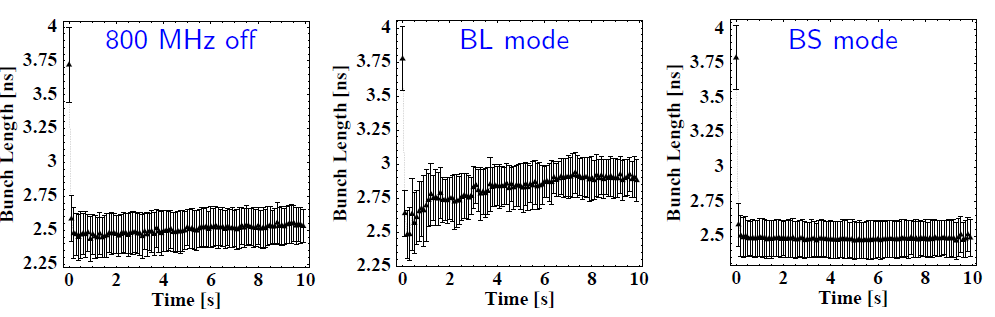}
	\caption{Measured average bunch length of the nominal LHC beam (average bunch intensity $1.25\times 10^{11}$) on the~26~GeV/c flat bottom of the CERN SPS as a function of time in a single 200~MHz rf system with $V_1 = 3$~MV (left) and in a double (200~MHz + 800~MHz) rf system with $V_4/V_1 = 0.23$ in the BL-mode (centre) and BS-mode (right). The bunch lengthening due to the coupled-bunch instability is suppressed only in the BS-mode. From Ref.~\cite{ES_PAC05}.}
	\label{fig:double_rf}
\end{figure}

The main applications and operation modes of a double rf system were described in the previous lecture\edt{~\cite{Lecture1}}. In the CERN SPS, the fourth harmonic rf system at 800~MHz is used in bunch-shortening (BS) mode during the whole acceleration cycle. In this mode the threshold of the coupled-bunch instability is increased by more than a factor \edt{of} five as compared to a single rf system, and the LHC-type beam can be delivered with the required bunch length and intensity. Note that only the BS-mode works for this application, see Fig.~\ref{fig:double_rf}.

\edt{The bunch-lengthening (BL) mode has a very high sensitivity to the phase errors between the two rf systems and a flat region in the synchrotron frequency distribution.}

\begin{figure}[htb]
	\centering
	\includegraphics[width = 0.5\textwidth]{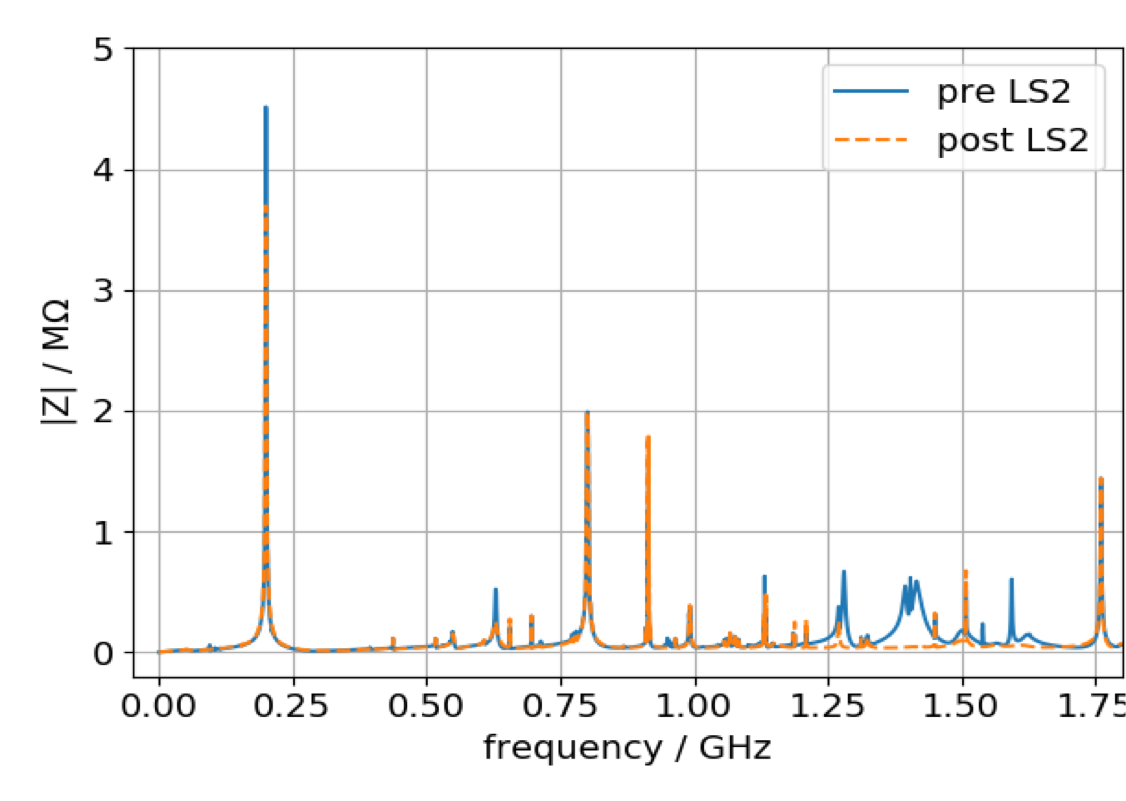}\\
	\begin{tabular}{cc}
	\includegraphics[width = 0.45\textwidth]{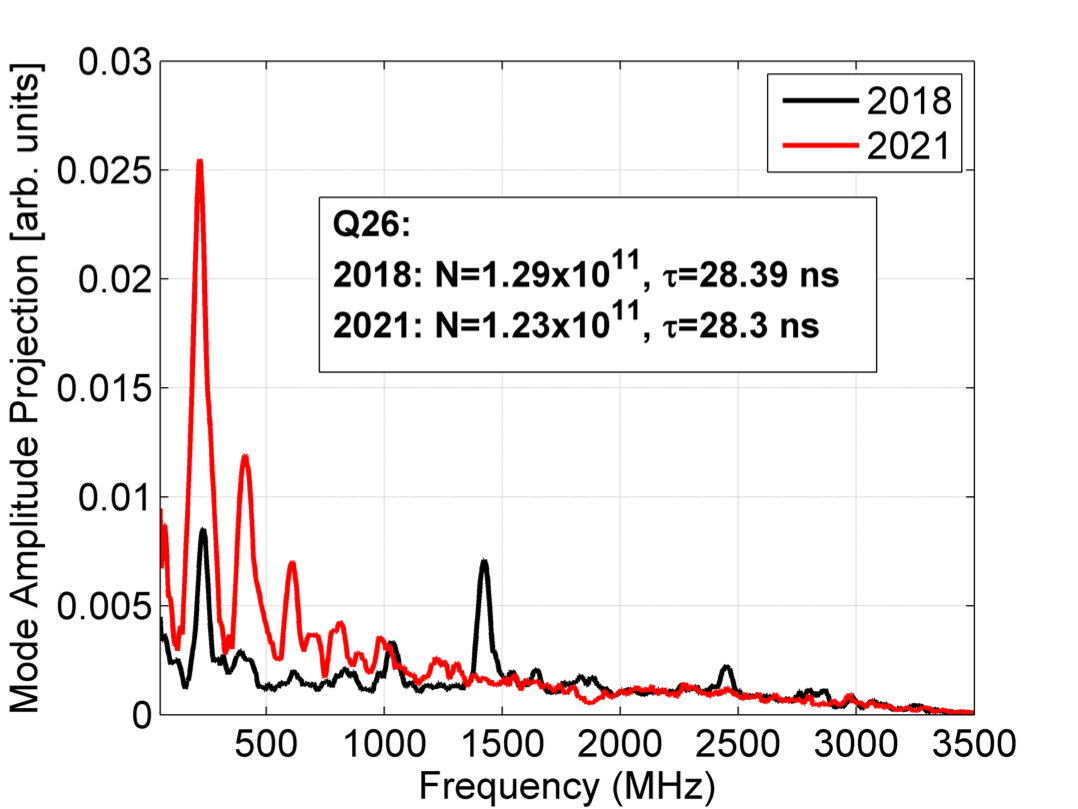} &
	\includegraphics[width = 0.45\textwidth]{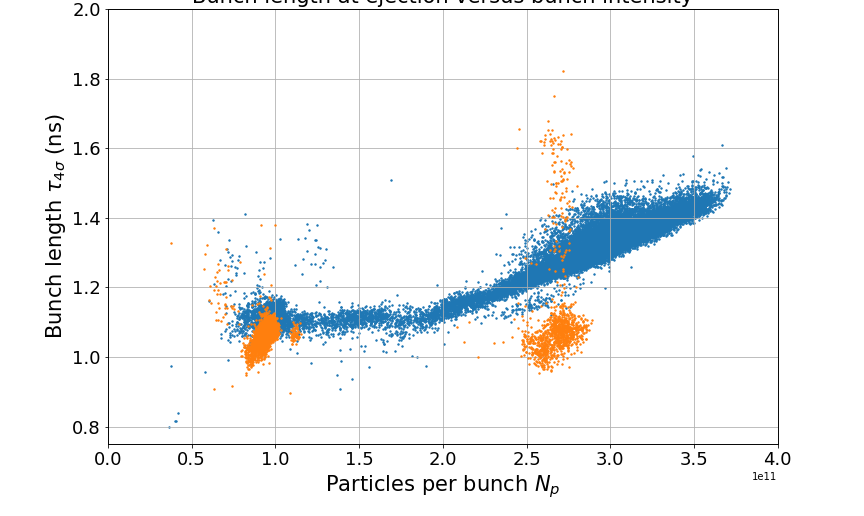}
	\end{tabular}
	\caption{Impedance reduction in the CERN SPS. Top: the SPS longitudinal impedance model before (blue) and after (orange) the long shutdown 2~\cite{Karpov2023, Argyropoulos2023}. Bottom left: comparison of the projected spectra of the unstable single bunches with intensity $N_p \approx 1.25\times 10^{11}$ injected into the SPS (Q26 optics, rf off) in 2018 (before) and 2021 (after the impedance reduction), from Ref.~\cite{Argyropoulos2023}. Bottom right: the measured bunch length at the SPS flat top (before extraction) as a function of bunch intensity in 2018 (blue) and 2021 (orange); the bunch lengthening above the~instability threshold, observed in 2018, is absent in 2021 for intensities up to $2.8\times 10^{11}$.}
	\label{fig:SPS_impedance_reduction}
\end{figure}

The passive cures attack the source of the instability, but often require significant efforts in \edt{the}~modification of the hardware.
\edt{The two impedance reduction campaigns in the CERN SPS, in 1999--2001 (with the shielding of the pumping ports) and in 2019--2020, were based} on the identification of the~impedance sources using the beam measurements \edt{method}~\edt{\cite{Bohl1997}} (Fig\edt{s}.~\ref{fig:unstable_spectrum}\edt{, \ref{fig:SPS_impedance_reduction}}). During the long shutdown 2 (2019--2020), the four long 200~MHz travelling wave cavities were replaced by six shorter structures (which allowed the peak voltage to be increased), the HOM damping at 623~MHz was improved and about 100 vacuum flanges were shielded, see Fig.~\ref{fig:SPS_impedance_reduction} (top).

As a result, the potential well distortion (bunch lengthening) is smaller, the threshold of the single-bunch instability at the SPS flat top is significantly increased (Fig.~\ref{fig:SPS_impedance_reduction}, bottom), and the HL-LHC beam with the required intensity can be stabilized~\cite{Argyropoulos2023}.

\section{\label{sec:conclusions} Conclusions}
In this lecture, the equations of longitudinal motion were extended to the case with intensity effects, when the total voltage seen by the particles contains, in addition to the rf voltage, the voltage induced by the beam in the accelerator impedance.
The stationary induced voltage leads to the potential well distortion. Its resistive part defines the synchronous phase shift. Its reactive part defines the shift of the~synchrotron frequency, with a strong dependence on the bunch length, and the bunch lengthening for a~given longitudinal emittance.

The linearized Vlasov equation was used to derive step-by-step the Lebedev matrix equation, which is a general equation for the harmonics of the perturbed line density and can be applied to any impedance model and rf system. This equation allowed \edt{us} to obtain an analytic expression for the LLD threshold of a single bunch in the presence of the constant inductive impedance. \edt{For a wide range of particle distributions described by the binomial family} this threshold vanishes, unless a cutoff frequency of the impedance is introduced. For the cutoff frequency $f_c \gg 1/\tau_\mathrm{full}$, the LLD threshold is inversely proportional to $f_c$ and the classical dependence on the bunch length in the fifth power is replaced by the~fourth.

The same Lebedev equation, applied to a multi-bunch beam with a narrow-band impedance, gives the threshold and the growth rates of the coupled-bunch instability, which are strongly dependent on the~bunch length and the particle distribution, as well as the spectrum of the unstable beam, used for \edt{the}~identification of the impedance sources. The Robinson instability, driven by the impedance at the~fundamental rf frequency, appears in the same formalism as a coupling of the two beam harmonics, corresponding to the upper and lower synchrotron sidebands. The broad-band impedance, responsible for the~loss of Landau damping, can significantly reduce the threshold of the coupled-bunch instability driven by the narrow-band impedance. The generalized threshold shows the key role of the LLD threshold and demonstrates how the two impedance contributions add up. These findings explain the beam observations in the CERN SPS and LHC and should be taken into account in the design of future accelerators.

Finally, the possible cures of the longitudinal beam instabilities were discussed with the examples of the higher harmonic rf system and of the impedance reduction campaign\edt{s} in the CERN SPS.

\section*{ACKNOWLEDGEMENTS}
The author thanks Ivan Karpov for his contribution \edt{to} the work described above, useful discussions and comments, as well as T.~Argyropoulos, H.~Damerau, A.~Lasheen and J.~Esteban M\edt{\"u}ller for their help in \edt{the} preparation of the lecture.

\bibliographystyle{unsrt}
\bibliography{refs}

@inproceedings{Tevatron,
      author        = {Moore, R. and Balbekov, V. I. and Jansson, A. and Lebedev, V. A. and Ng, K. Y. and Shiltsev, V. D. and Tan, C. Y.},
      title         = "{Longitudinal bunch dynamics in the Tevatron}",
      pages         = "1751--1753",
      year          = "2003",
      url           = "https://accelconf.web.cern.ch/p03/PAPERS/TPPB066.pdf",
      booktitle = "{Proceedings of the 2003 Particle Accelerator Conference}",
      address = "{Portland, Oregon, USA}",
}

@inproceedings{RHIC,
  author={M. {Blaskiewicz} and J. M. {Brennan} and P. {Cameron} and W. {Fischer} and J. {Wei} and A. {Luque} and H. {Schamel}},
  booktitle={Proceedings of the 2003 Particle Accelerator Conference}, 
  title="{Longitudinal solitons in RHIC}", 
  year={2003},
  number={},
  pages={3029--3031},
  doi={10.1109/PAC.2003.1289803},
  address = {Portland, Oregon, USA},
  }

@inproceedings{SPS,
    author = "Shaposhnikova, E.",
    title = "{Cures for beam instabilities in the CERN SPS and their limitations}",
    booktitle = "{39th ICFA Advanced Beam Dynamics Workshop on High Intensity High Brightness Hadron Beams 2006}",
    pages = "153--155",
    year = "2006",
    address = "Tsukuba, Japan",
    url = "http://accelconf.web.cern.ch/abdwhb06/PAPERS/TUBX05.PDF",
}

@inproceedings{LHC,
  title="{Loss of Landau damping in the LHC}",
  author={E. {Shaposhnikova} and T. {Argyropoulos} and T. {Bohl} and C. {Bhat} and P. {Baudrenghien} and A. {Butterworth} and T. {Mastoridis} and J. {Muller} and G. {Papotti} and J. {Tuckmantel} and W. V. {Delsolaro}},
  year={2011},
  booktitle = "{Proceedings of 2nd International Particle Accelerator Conference}",
  url = "https://accelconf.web.cern.ch/IPAC2011/papers/MOPC057.pdf",
  address = "San Sebasti\'an, Spain",
  pages = "211--214",
}

@phdthesis{JMPhD,
      author        = "Esteban M{\"u}ller, J. F.",
      title         = "{Longitudinal intensity effects in the CERN Large Hadron Collider}",
      type          = "{Ph.D. Thesis}",
      reportNumber  = "CERN-THESIS-2016-066",
      url           = "https://cds.cern.ch/record/2196930",
      year = "2016",
      school        = "EPFL, Lausanne, Switzerland"
}

@book{NG,
author = {Ng, K. Y.},
title = {Physics of Intensity Dependent Beam Instabilities},
publisher = {World Scientific, Singapore},
year = {2006},
doi = {10.1142/5835},
address = {},
edition   = {},
}

@article{Lebedev1968, 
    author={A. N. {Lebedev}}, 
    journal={Atomic Energy}, 
    title={Coherent synchrotron oscillations in the presence of a space charge}, 
    year={1968}, 
    volume={25}, 
    number={2}, 
    pages={851--856}, 
    doi={10.1007/BF01121037}, 
}

@article{Balbekov-Ivanov1986,
    author={V. I. {Balbekov} and S. V. {Ivanov}},
    title= "{Longitudinal instability threshold of bunched beams in proton synchrotrons}",
    journal={Atomic Energy}, 
    volume={60},
    number={1},
    pages={58--66},
    year={1986},
    doi = {10.1007/BF01129839}
}

@article{Hofmann1979, 
    author={A. {Hofmann} and F. {Pedersen}}, 
    journal={IEEE Transactions on Nuclear Science}, 
    title={Bunches with Local Elliptic Energy Distributions}, 
    year={1979}, 
    volume={26}, 
    number={3}, 
    pages={3526--3528}, 
    doi={10.1109/TNS.1979.4330088}, 
}

@article{YHChin1983,
    author = "Chin, Y. H. and Satoh, K. and Yokoya, K.",
    title = "{Instability of a bunched beam with synchrotron frequency spread}",
    journal = "Particle Accelerators",
    volume = "13",
    pages = "45--66",
    year = "1983",
    url = "https://s3.cern.ch/inspire-prod-files-2/2867431f77457ceca5bdedc874763555",
}

@article{vKampen1,
title = "On the theory of stationary waves in plasmas",
journal = "Physica",
volume = "21",
number = "6",
pages = "949--963",
year = "1955",
issn = "0031--8914",
doi = "https://doi.org/10.1016/S0031-8914(55)93068-8",
url = "http://www.sciencedirect.com/science/article/pii/S0031891455930688",
author = "N. G. {Van Kampen}",
}

@article{vKampen2,
title = "The dispersion equation for plasma waves",
journal = "Physica",
volume = "23",
number = "6",
pages = "641--650",
year = "1957",
issn = "0031--8914",
doi = "https://doi.org/10.1016/S0031-8914(57)93718-7",
url = "http://www.sciencedirect.com/science/article/pii/S0031891457937187",
author = "N. G. {Van Kampen}",
}

@inproceedings{burov2012van,
    title="{Van Kampen modes for bunch longitudinal motion}",
    author={A. {Burov}},
    booktitle = "{Proceedings of 46th ICFA Advanced Beam Dynamics Workshop on High-Intensity and High-Brightness Hadron Beams}",
    address = {Morschach, Switzerland},
    year={2010},
    url = {http://accelconf.web.cern.ch/HB2010/papers/tuo1c03.pdf},
    eprint={1207.5826},
    archivePrefix={arXiv},
    primaryClass={physics.acc-ph}
}

@techreport{Oide_Yokoya1990,
    author = {Oide, K. and Yokoya, K.},
    title = "{Longitudinal single bunch instability in electron storage rings}",
    year = {1990},
    institution={KEK},
    number={KEK-Preprint-90-10},
    url = {http://cds.cern.ch/record/208816/files/kek-preprint-90-10.pdf},
}

@misc{MELODY,
author = {Karpov, I.},
title = {{MELODY}: Matrix Equations for {LO}ngitudinal beam {DY}namics},
note = {\newline\url{https://gitlab.cern.ch/ikarpov/melody}},
}

@techreport{Shaposhnikova1994,
      author        = "Shaposhnikova, E.",
      title         = "{Bunched beam transfer matrices in single and double rf systems}",
      institution   = "CERN",
      address       = "Geneva, Switzerland",
      number        = "CERN-SL-94-19-RF",
      year          = "1994",
      url           = "https://cds.cern.ch/record/267121",
}

@article{Dyachkov,
    author = "M. {D'yachkov} and R. {Baartman}",
    title = "{Longitudinal single bunch stability}",
    journal = "Particle Accelerators",
    volume = "50",
    pages = "105--116",
    year = "1995",
    url = "https://cds.cern.ch/record/1120215/files/p105.pdf",
}

@misc{BLOND2020,
note = "{CERN Beam Longitudinal Dynamics code BLonD, \url{http://blond.web.cern.ch}}",
}

@article{CYTanABurov2012,
  title = "{Phase modulation of the bucket stops bunch oscillations at the Fermilab Tevatron}",
  author = {Tan, C. Y. and Burov, A.},
  journal = {Phys. Rev. ST Accel. Beams},
  volume = {15},
  issue = {4},
  pages = {044401},
  numpages = {9},
  year = {2012},
  publisher = {American Physical Society},
  doi = {10.1103/PhysRevSTAB.15.044401},
  url = {https://link.aps.org/doi/10.1103/PhysRevSTAB.15.044401}
}

@inproceedings{Shaposhnikova2014,
  title="{Flat bunches in the LHC}",
  author={E. {Shaposhnikova} and T. {Argyropoulos} and P. {Baudrenghien} and J. F. {Esteban M\"uller} and T. {Mastoridis} and G. {Papotti} and B. {Salvant} and H. {Timko} and C. M. {Bhat} and A. V. {Burov}},
  year={2014},
  booktitle = "{Proceedings of 5th International Particle Accelerator Conference}",
  url = "http://accelconf.web.cern.ch/IPAC2014/papers/tupme028.pdf",
  address = "Dresden, Germany",
  pages = "1413--1415",
}

@book{Chao,
      author        = "Chao, A. W.",
      title         = "{Physics of collective beam instabilities in high energy accelerators}",
      publisher     = "Wiley",
      address       = "New York, NY",
      year          = "1993",
      url           = "http://cds.cern.ch/record/246480",
      pages = {279--305},
}

@article{Ivan,
  title = {Thresholds for loss of {L}andau damping in longitudinal plane},
  author = {Karpov, I. and Argyropoulos, T. and Shaposhnikova, E.},
  journal = {Phys. Rev. Accel. Beams},
  volume = {24},
  pages = {011002},
  year = {2021},
  doi = {10.1103/PhysRevAccelBeams.24.011002},
}

@article{Karpov2024,
  title = {Generalized threshold of longitudinal multibunch instability in synchrotrons},
  author = {Karpov, I. and Shaposhnikova, E.},
  journal = {Phys. Rev. Accel. Beams},
  volume = {27},
  pages = {074401},
  year = {2024},
  doi = {10.1103/PhysRevAccelBeams.27.074401},
}

@article{Karpov2023,
  title = {Longitudinal mode-coupling instabilities of proton bunches in the {CERN} {S}uper {P}roton {S}ynchrotron},
  author = {Karpov, I.},
  journal = {Phys. Rev. Accel. Beams},
  volume = {26},
  pages = {014401},
  year = {2023},
  doi = {10.1103/PhysRevAccelBeams.26.014401},
}

@inproceedings{Karpov2019,
  title = {Longitudinal coupled-bunch instability evaluation for {FCC-hh}},
  author = {Karpov, I. and Shaposhnikova, E.},
  booktitle = {Proceedings of the 10th International Particle Accelerator Conference (IPAC'19)},
  address = {Melbourne, Australia},
  year = {2019},
  pages = {297--300},
}

@inproceedings{Argyropoulos2023,
  title = {Experimental confirmation of the impedance reduction campaign in the {CERN} {SPS}},
  author = {Argyropoulos, T. and Damerau, H. and Lasheen, A. and Papotti, G. and Karpov, I.},
  booktitle = {Proceedings of the 14th International Particle Accelerator Conference (IPAC'23)},
  address = {Venice, Italy},
  year = {2023},
}

@article{Bohl1997,
  title = {Measuring the resonance structure of accelerator impedance with single bunches},
  author = {Bohl, T. and Linnecar, T. P. R. and Shaposhnikova, E.},
  journal = {Phys. Rev. Lett.},
  volume = {78},
  pages = {3109--3112},
  year = {1997},
  doi = {10.1103/PhysRevLett.78.3109},
}

@article{Lasheen2017,
  title = {Evaluation of the {CERN} {S}uper {P}roton {S}ynchrotron longitudinal impedance from measurements of the quadrupole frequency shift},
  author = {Lasheen, A. and Shaposhnikova, E.},
  journal = {Phys. Rev. Accel. Beams},
  volume = {20},
  pages = {064401},
  year = {2017},
  doi = {10.1103/PhysRevAccelBeams.20.064401},
}

@article{Shaposhnikova1999,
  title = {Analysis of coupled bunch instability spectra},
  author = {Shaposhnikova, E.},
  journal = {AIP Conf. Proc.},
  volume = {496},
  pages = {256},
  year = {1999},
}

@article{Haissinski,
  title = {Exact longitudinal equilibrium distribution of stored electrons in the presence of self-fields},
  author = {Haissinski, J.},
  journal = {Il Nuovo Cimento B},
  volume = {18},
  pages = {72--82},
  year = {1973},
}

@techreport{Robinson1964,
  title = {Stability of beam in radiofrequency system},
  author = {Robinson, K. W.},
  institution = {Cambridge Electron Accelerator},
  address = {Cambridge, MA, USA},
  number = {CEAL-1010},
  year = {1964},
}

@article{Pedersen1975,
  title = {Beam loading effects in the {CERN} {PS} booster},
  author = {Pedersen, F.},
  journal = {IEEE Trans. Nucl. Sci.},
  volume = {22},
  number = {3},
  pages = {1906--1909},
  year = {1975},
}

@article{Lasheen2018,
  title = {Beam measurement of the high frequency impedance sources with long bunches in the {CERN} {S}uper {P}roton {S}ynchrotron},
  author = {Lasheen, A. and Shaposhnikova, E.},
  journal = {Phys. Rev. Accel. Beams},
  volume = {21},
  pages = {034401},
  year = {2018},
  doi = {10.1103/PhysRevAccelBeams.21.034401},
}

@misc{SPSimp,
  note = {{CERN SPS} longitudinal impedance model, 
  \newline
  \url{https://gitlab.cern.ch/longitudinal-impedance/SPS}},
}

@article{EstebanMuller2015,
  title = {High-accuracy diagnostic tool for electron cloud observation in the {LHC} based on synchronous phase measurements},
  author = {Esteban M\"uller, J. F. and Baudrenghien, P. and Mastoridis, T. and Shaposhnikova, E. and Valuch, D.},
  journal = {Phys. Rev. ST Accel. Beams},
  volume = {18},
  pages = {112801},
  year = {2015},
  doi = {10.1103/PhysRevSTAB.18.112801},
}

@inproceedings{ES_PAC05,
  title = {Beam transfer functions and beam stabilisation in a double rf system},
  author = {Shaposhnikova, E. and Bohl, T. and Linnecar, T.},
  booktitle = {Proceedings of the 2005 Particle Accelerator Conference (PAC'05)},
  address = {Knoxville, TN, USA},
  year = {2005},
  pages = {2300--2302},
}

@inproceedings{BohlEPAC04,
  author       = {Bohl, T. and Linnecar, T. and Shaposhnikova, E. and T{\"u}ckmantel, J. and Hofmann, A.},
  title        = {Energy loss of a single bunch in the {CERN} {SPS}},
  booktitle    = {Proc. EPAC'04, Lucerne, Switzerland},
  year         = {2004},
  note         = {paper WEPLT036}
}

@inproceedings{Schwarz2020,
  author       = {Schwarz, M. and Farricker, A. and Karpov, I. and Lasheen, A.},
  title        = {Synchronous phase shift measurements for evaluation of the longitudinal impedance model at the {CERN} {SPS}},
  booktitle    = {Proc. ICFA mini-Workshop on Mitigation of Coherent Beam Instabilities in Particle Accelerators (MCBI 2019), Zermatt, Switzerland, CERN Yellow Rep. Conf. Proc. 9},
  pages        = {318},
  year         = {2020}
}

@inproceedings{Hofmann77,
  author       = {Hofmann, A.},
  title        = {Single-beam collective phenomena --- longitudinal},
  booktitle    = {Proc. First Course Int. School of Particle Accelerators, Erice, 1976, CERN Report 77-13},
  pages        = {139},
  year         = {1977}
}

@misc{Lecture1,
  author       = {Shaposhnikova, E.},
  title        = {Longitudinal beam dynamics {II}. {Lecture} 1: single and double rf systems},
  note         = {These proceedings},
  year         = {}
}

\end{document}